\documentclass{article} 

\usepackage{fullpage} 
\usepackage{stmaryrd}  

\usepackage{mathptmx}
\usepackage{verbatim}
\usepackage{amssymb}  
\usepackage{epsfig}

\usepackage{amsthm} 
\newtheorem{definition}{Definition}[section]
\newtheorem{example}{Example}[section]
\newtheorem{theorem}{Theorem}[section]

\usepackage{url}
\usepackage{amssymb}

\usepackage{epsfig}
\usepackage{wrapfig}
\usepackage{subfigure}
\usepackage{rotating}
\usepackage{moreverb}
\usepackage{fancyvrb}

\def\bd{\begin{description}}
\def\ed{\end{description}}
\def\bc{\begin{center}}
\def\ec{\end{center}}
\def\bi{\begin{itemize}}
\def\ei{\end{itemize}}
\def\be{\begin{enumerate}}
\def\ee{\end{enumerate}}
\def\ba{\begin{array}}
\def\ea{\end{array}}

\newcommand{\chr}{{CHR}}

\newcommand{\mcP}{\ensuremath{\mathcal{P}}}

\newcommand{\bbB}{\ensuremath{\mathbb{B}}}

\newcommand{\bbG}{\ensuremath{\mathbb{G}}}

\newcommand{\bbV}{\ensuremath{\mathbb{V}}}

\def\tuple#1{\langle #1 \rangle}

\newcommand{\der}{\ensuremath{\rightarrowtail}}

\newcommand{\act}[3]{#1{\#}#2{:}#3}

\newcommand{\atsign}{:}  

\newcommand{\ELsimparrow}[0]{<=>}

\newcommand{\bda}{\[\ba}
\newcommand{\eda}{\ea\]}

\newcommand{\arat}{\ar@}

\newcommand{\ELpartransstar}{\rightarrowtail^*_{\mid\mid}}

\newcommand{\ELstcup}{\cup}
\newcommand{\ELgoaltranssf}[1]{\stackrel{#1}{\rightarrowtail}}
\newcommand{\ELpartranssf}[1]{\stackrel{#1}{\rightarrowtail_{\mid\mid}}}

\newcommand{\ELchrstate}[2]{\langle #1\mid#2 \rangle}

\newcommand{\ELsgap}{\quad}

\newcommand{\ELtlabel}[1]{\mbox{{\bf(#1)}}}

\newcommand{\ELfig}[3]
        {\begin{figure}[htb]
\rule{\textwidth}{0.5pt}#3\
\rule{\textwidth}{0.5pt}
        \caption{\label{#1}#2}
\end{figure}}

\newcommand{\ELmyirule}[2]{{\renewcommand{\arraystretch}{1.2}\ba{c} #1
                      \\ \hline #2 \ea}}

\newcommand{\chrmp}{CHRmp}
\newcommand{\chrd}{CHRd}
\newcommand{\chre}{CHRe}
\newcommand{\chrt}{CHRt}

\newcommand{\myparagraph}[1]{\textbf{#1.}}

\newcommand{\parmapsto}{\Mapsto}  

\begin{document}

  \title
        {Parallelism, Concurrency and Distribution in Constraint Handling Rules: A Survey}  

  \author
         {THOM FR\"{U}HWIRTH\\
         University of Ulm, Germany
         }

\label{firstpage}

\maketitle

\begin{abstract}

Constraint Handling Rules (CHR) is both an effective concurrent declarative programming language and a versatile computational logic formalism. 
In CHR, guarded reactive rules rewrite a multiset of constraints. Concurrency is inherent, since rules can be applied to constraints in parallel.

In this comprehensive survey, we give an overview of concurrent, parallel as well as distributed CHR semantics, standard and more exotic, 
that have been proposed over the years at various levels of refinement. 
These semantics range from the abstract to the concrete. They are related by formal soundness results.
Their correctness is proven as a correspondence between parallel and sequential computations.

On the more practical side, we present common concise example CHR programs that have been widely used in experiments and benchmarks. 
We review parallel and distributed CHR implementations in software as well as hardware. 
The experimental results obtained show a parallel speed-up for unmodified sequential CHR programs. 
The software implementations are available online for free download and we give the web links.

Due to its high level of abstraction, the CHR formalism can also be used to implement and analyse models for concurrency. 
To this end, the Software Transaction Model, the Actor Model, Colored Petri Nets 
and the Join-Calculus have been faithfully encoded in CHR.
Finally, we identify and discuss commonalities of the approaches surveyed and indicate what problems are left open for future research.

Under consideration in Theory and Practice of Logic Programming (TPLP).
  \end{abstract}


Declarative Programming,
Concurrent Constraint Programming,

Semantics, Rewriting, Concurrency Models.

\tableofcontents  

\section{Introduction}
\label{sec:intro}

Parallelism has become an eminent topic in computer science again with the widespread arrival of multi-core processors.
With the proliferation of mobile devices and the promises of the internet-of-things, distribution is another major topic, intertwined with parallelism.
Parallel and distributed programming is known to be difficult. Declarative programming languages promise to ease the pain.
This survey shows how parallelism and distribution are addressed in the declarative language Constraint Handling Rules.

\myparagraph{Basic Notions}
Before we start with our survey, we shortly clarify the essential concepts at stake and introduce Constraint Handling Rules.
The technical terms of concurrency, parallelism and distribution have an overlapping meaning, and
processes are another central notion in this context. Due to their generality they are hard to define precisely:
\begin{description}
\item {\em Concurrency}\index{concurrency} allows for logically more or less independent computations, be they sequential or parallel. This abstract concept thus supports the modular design of independent program components that can be composed together.

\item {\em Parallelism}\index{parallelism} allows for computations that happen simultaneously, at the same time, thus hopefully improving performance. 
On the downside, sequential programs usually have to be rewritten to be able to run in parallel.
With the arrival of multi-core processors, it has become a dominant computation model.
The processors may have access to a shared memory to exchange information.

\item {\em Distribution} allows for program components that are located on physically distributed decentralized networked processors. 
Each processor has its own local memory (distributed memory).
Personal computers, the internet and mobile devices have enforced this computational paradigm.
Distribution introduces modularity and potential parallelism, but also the need for communication between the components.

\item {\em Processes}\index{process} 
are programs that are executed independently but can interact with each other.  
Processes can either execute local actions\index{action} or {communicate}, coordinate
  and {synchronize} by passing (sending and receiving) messages.  
Depending on context and level of abstraction, processes are also called threads, workers, tasks, activities or even agents.
\end{description}
Concurrency and distribution are easier with declarative programming
languages, since they are compositional:
different computations can be composed into one without unintended interference.
Moreover, declarative languages offer a wealth of program analysis and reasoning techniques. 

\myparagraph{Constraint Handling Rules (CHR)} 
CHR is both an effective concurrent declarative con\-straint-based programming language and a versatile computational logic formalism~\cite{fru_chr_book_2009,chr_survey_tplp10,chr-thesis-book,fruhwirth2015constraint,chrwebsite}. 
CHR has its roots in constraint logic programming and concurrent constraint programming, but also integrates ideas from multiset transformation and rewriting systems. 
While conceptually simple, 
CHR is distinguished by a remarkable combination of desirable features:
\begin{itemize}
\item a semantic foundation in classical logic as well as in linear logic \cite{betz2014unified}, 
\item an effective and efficient sequential and parallel execution model \cite{chr-thesis-book},
\item a proof that every algorithm can be expressed with best known time and space complexity \cite{sney_schr_demoen_chr_complexity_toplas09},
\item up to a million rule applications per second due to CHRs novel rule execution strategy based on lazy matching without conflict resolution \cite{vanweert_lazy_evaluation_tkde10},
\item guaranteed properties like the anytime algorithm and online algorithm properties \cite{Abdennadher+99},
\item program analysis methods for deciding essential properties like confluence and program equivalence \cite{abd_fru_equivalence_cp99}.
\end{itemize}
The given references are meant to serve as starting points into the respective themes. 
One could continue with their references but also the papers that reference them.

Information on CHR can be found online at  
\url{http://www.constraint-handling-rules.org}, 
including news, tutorials, papers, bibliography, online demos and free downloads of the language.

\myparagraph{Minimum Example}
Assume we would like to compute the minimum of some numbers, 
given as multiset {\tt min($n_1$), min($n_2$),\ldots, min($n_k$)}.
We interpret the constraint (predicate) {\tt min($n_i$)} to mean that the number $n_i$ is a candidate
for the minimum value.
We make use of the following CHR rule that filters the candidates.
\begin{verbatim}
min(N) \ min(M) <=> N=<M | true.
\end{verbatim}
The rule consists of a left-hand side, on which a pair of constraints has to be matched,
a guard check {\tt N=<M} that has to be satisfied, and an empty right-hand side denoted by {\tt true}.
In effect, the rule takes two {\tt min} candidates and removes the one with the
larger value (constraints after the \verb+\+ symbol are to be removed). 
Starting with a given initial state, CHR rules are applied exhaustively, resulting in a final state.
Note that CHR is a committed-choice language, i.e. there is no backtracking in the rule applications. 
Here the rule keeps on going until only one, thus the smallest
value, remains as single {\tt min} constraint.
Note that the {\tt min} constraints behave both as operations (removing other constraints) and as data (being removed).
This abstraction is characteristic of the notion of constraint.

A {\em state} is a multiset of constraints.
In a sequential computation, we apply one rule at a time to a given state.
A possible computation sequence is (where we underline constraints involved
in a rule application):
\begin{center}
\noindent {\tt \underline{min(1)}, \underline{min(0)}, min(2), min(1) $\mapsto$}\\
{\tt \underline{min(0)}, \underline{min(2)}, min(1) $\mapsto$}\\
{\tt \underline{min(0)}, \underline{min(1)} $\mapsto$}\\
{\tt {min(0)}}
\end{center}
The final state is called {\em answer}. The remaining constraint contains the minimum value, in this case zero.

By the way, CHR insists on multisets so one can directly model resources as constraints, for example:
\begin{verbatim}
buy : cup \ euro, euro <=> coffee.
\end{verbatim}
This rule expresses that we get a coffee for two euros if we have a cup.
As we will see, there are also some semantics and implementations of CHR that are set-based.

\myparagraph{Concurrency and Parallelism in CHR}
One of the main features of CHR is its inherent concurrency.
Intuitively, in a parallel execution of a CHR program, rules can be applied to
separate parts of a {state} in parallel. 
As we will see, CHR rules can even be applied in parallel to overlapping parts of a state,
in principle without the need to change the program. 
This is referred to as \index{logical parallelism}{\em logical parallelism} or 
\index{declarative concurrency}{\em declarative concurrency}.

The rule of {\tt min} can be applied in parallel to different parts of the state:
\begin{center}
{\tt \underline{min(1)}, \underline{min(0)}, ~~~~~ \underline{min(2)}, \underline{min(1)} $\mapsto$}\\
{\tt \underline{min(0)}, ~~~~~~~~~~~~~ \underline{min(1)} $\mapsto$}\\
{\tt {min(0)}}\\
\end{center}
We arrive at the answer in less computation steps than with the sequential execution.

The rule can also be applied in parallel to
overlapping parts of the state, provided the overlap is not removed by any rule. 
For example, let the overlap be the constraint {\tt min(0)}.
Then the three pairs {\tt min(0), min(1)}, {\tt min(0), min(1)} and {\tt min(0), min(2)}
can be matched to different rule instances. 
(Note that we always match the same {\tt min(0)}, but that we have two copies of {\tt min(1)}.)
These rules can be applied at the same time, since the common (overlapping) constraint {\tt min(0)} is not removed.
\begin{center}
{\tt \underline{\underline{\underline{min(0)}}},~~ \underline{min(1)}, \underline{min(2)}, \underline{min(1)} $\mapsto$}\\
{\tt {min(0)}}
\end{center}
So this is another, even shorter way to arrive at the same answer.

In CHR, concurrently executing processes are CHR constraints that communicate
via a shared built-in constraint {store}.
The built-in constraints take the role of (partial) messages and variables take
the role of communication channels.  

\myparagraph{Guaranteed Properties of CHR}
First of all, the essential {\em monotonicity property} of CHR 
means that adding constraints to a state cannot
inhibit the applicability of a rule. 
(Rule matching and guards check for presence of certain constraints, never absence.)
Among other things, this monotonicity enables decidable program analyses
and helps declarative concurrency.
Most, but not all semantics that we introduce enjoy the monotonicity property.

Now assume that while the program runs, we
add another constraint. It will eventually participate in the
computation in that a rule will be applied to it. The answer will be
as if the newly added constraint had been there from the beginning but ignored
for some time.
This property of a CHR program is called {\em incrementality} or 
\index{online algorithm}{\em online algorithm property} 
and directly follows from monotonicity.

Furthermore, in CHR, we can stop the computation at {any time} and observe 
the current state as intermediate answer. We can then continue by applying
rules to this state without the need to recompute from scratch. If
we stop again, we will observe a next intermediate answer that is closer to the
final answer. 
This property of a CHR program is called the
\index{anytime algorithm}{\em anytime algorithm property}.
Note that by this description, an anytime algorithm is also an
\index{approximation algorithm}{\em approximation algorithm},
since intermediate answers more and more approximate the final answer.

\myparagraph{Desirable Property of Confluence} 
This property of a program guarantees that any computation starting from a given initial state results in the same answer no matter which of the applicable rules are applied. There is a decidable, sufficient and necessary syntactic condition to analyse confluence of terminating programs and to detect rule pairs that lead to non-confluence when applied. 
Among other things, confluence implies that rules can be applied in parallel, with the same result as any sequential computation,
without the need for any modification of the given program. If on the other hand a program is not confluent, it may have to be rewritten to ensure proper parallel execution. This rewriting is aided by the method of completion, which automatically adds rules to a program to make it confluent
(but may not terminate).

An introduction into all these properties can be found in~\cite{fru_chr_book_2009}.
In the next section we will discuss desirable properties that characterize the correspondence between different semantics of CHR.

\myparagraph{Overview of the Survey and its Structure}
The richness of topics in this survey, from formal semantics to hardware implementation and more, poses a challenge for the structure of this text. We decided to go from abstract to concrete while making sure concepts are introduced in sections before they are referred to in later sections.
\begin{description}
\item {\em Section 2-4: Abstract Parallel CHR Semantics, Example Programs, Extension by Transactions.}
In the next section we define abstract syntax and abstract operational semantics for CHR.
One sequential transition describes rule applications, another one parallel transitions, 
a trivial third one connects the two. 
The essential correctness properties of monotonicity, soundness and serializability are introduced.
In Section 3, we present common classic CHR example programs based on well-known algorithms. 
Often one rule suffices. All but one of the programs can be run in parallel without change.
In Section 4, we extend abstract parallel CHR with transactions, a popular and essential concept in concurrency.

\item {\em Section 5-6: Refining the Parallel Semantics and its Implementation.}
In Section 5, we refine our abstract semantics by differentiating between a goal and a constraint store.
The goal holds active constraints to execute them as processes in operation, the constraint store holds inactive constraints as data.
This implies that we now have to account for the in-activation (suspension) and re-activation (wake-up) of user-defined constraints.
In Section 6, we describe an implementation of the refined semantics in Haskell using software transactions and 
the result of benchmark experiments showing parallel speed-ups.

\item {\em Section 7-8: Excursion: Set-Based Massive Parallelism and Hardware Implementations.}
Section 7 introduces a more exotic abstract semantics that is massively parallel. 
It is also set-based.
This theoretical model in the extreme case allows to find primes in constant time and 
to solve SAT problems in linear time. This comes with a cost: soundness only holds under a certain condition.
We then move on to more mundane fast hardware implementations of the parallel CHR semantics introduced in Section 8 and again present some experimental evidence.
It is typically one order of magnitude faster than the fastest software implementations.
The translation scheme of the hardware implementations also applies to procedural languages like C and Java.

\item {\em Section 9: Distribution in CHR.}
In Section 9 we discuss two distributed semantics for CHR, where the constraint store and computations are decentralized by introducing the notion of locations.
Distribution requires a syntactic restriction on CHRs rule heads to ensure shared variables as communication channels among locations.
The first semantics is informal and set-based, the second one full-fledged. Both semantics allow for propagation rules.
Both semantics have been implemented.

\item {\em Section 10: Concurrency Models in CHR.}
Last but not least, in Section 10 we shortly show the high-level encoding common formal models of concurrency in CHR on four concrete models:
the Software Transaction Model, the Actor Model, Colored Petri Nets 
and the Join-Calculus have been faithfully embedded in CHR 
to enable comparison and further investigation by the program analyses available in CHR.
The embeddings have been proven correct. Some embeddings are available online.

\item {\em Section 11-12: Discussion and Conclusions.}
We end the paper with a discussion, directions for future work and in Section 11 with conclusions.
\end{description}

Within the sections, we also try to follow a standard structuring where applicable:
We define the parallel or distributed semantics at hand and discuss its correspondence to the standard sequential CHR semantics.
This usually done by proving the properties of soundness and serializability, which are notions of correctness.
Another property of interest is monotonicity, which is also enjoyed by standard CHR.
For software and hardware implementations, we give free download links and we summarize experimental results found in the literature.
We illustrate the approaches to semantics and implementation with additional examples.

For a better reading experience, we use the editorial we throughout. 
Of course it refers to different authors in different sections of this paper.

\section{Parallel Abstract Operational Semantics of CHR}
\label{sec:chr-intro}

We will present the sequential equivalence-based abstract CHR semantics 
and extend it with parallelism. 
We just need a sequential transition describes rule applications, another one parallel transitions, a trivial third one that connects the two.
We also introduce the three properties that prove the correctness of a given semantics with regard to a more abstract or a sequential semantics: 
monotonicity, soundness and serializability.
We assume basic familiarity with first-order predicate logic and state transition systems.
Readers familiar with CHR can skip most of this section.
We start with some preliminaries.

\subsection{Semantics of CHR and their Properties}

{\em Structural Operational Semantics (SOS)} is a common inductive approach to describe the behavior of programming languages, in particular concurrent ones.
In SOS, a {\em state transition system} specifies the computations. Transitions rewrite states and take the form of inference rules. All semantics of CHR, sequential or parallel, employ this approach.

\myparagraph{Semantics for sequential CHR} 
They exist in various formulations and at various levels of refinement,
going from the abstract to the concrete (refined)~\cite{fru_chr_book_2009,betz_raiser_fru_execution_model_iclp10}:
\begin{itemize}
\item The {\em very abstract semantics} \cite{fru_chr_book_2009} is close to modus ponens of predicate logic. 
\item The {\em abstract semantics} \cite{Abdennadher+99} is the classical basis for CHR program analysis
and its properties. 
\item The more recent {\em state-equivalence-based abstract semantics} \cite{raiser_betz_fru_equivalence_revisited_chr09} 
will be the starting point of our survey.
We will extend it with parallelism.
\item The {\em refined semantics} \cite{duck_stuck_garc_holz_refined_op_sem_iclp04} describes more concretely the actual behavior of CHR implementations.
All more concrete parallel semantics of CHR are based on it. 
\end{itemize}
In addition, several alternative operational semantics for sequential CHR have been proposed.

\myparagraph{Soundness and Serializability}
The {\em correctness} of a more refined semantics is shown by its {\em soundness} with regard to a more abstract semantics.
This means that for each computation in the refined semantics, there is a corresponding computation in the abstract semantics. The converse (completeness) typically does not hold, because refined semantics are more concrete and thus rule out certain computations.
When we introduce a parallel semantics for CHR, it will be related by soundness to a more abstract semantics and/or the sequential part of the semantics.

Actually, the {\em interleaving semantics} approach to concurrency is defined by the fact that for
each possible parallel computation, there exists a corresponding sequential computation with
the same result.
The sequential computation uses interleaving of the different
parallel computations.
This means that a parallel computation step can be simulated
by a sequence of sequential computation steps.
This correspondence property is called {\em serializability (sequential consistency)}.
Most semantics we discuss are correct in this way.

\subsection{Abstract Syntax of CHR}

{\em Constraints} are relations, distinguished predicates of first-order predicate logic.
We differentiate between 
two kinds of constraints: {\em built-in (pre-defined) constraints} 
and
{\em user-defined (CHR) constraints} 
which are defined by the rules in a CHR program.
Built-in constraints can be used as tests in the guard as well as for auxiliary computations in the body of a rule.
In this survey, besides the trivial constraint {\tt true}, we will have syntactical equality {\tt =} between logical terms and equations between arithmetic expressions.

\begin{definition}
{\rm

A {\em goal} is a conjunction of built-in and user-defined constraints.
A {\em state} is also a goal.
Conjunctions are understood as {\em multisets} of their conjuncts.
We will use letters such as $A,B,C,D,E,\ldots$ for goals and $S$ and $T$ for states.

A {\em \chr\ program} is a finite set of rules.  
A {\em (generalized) simpagation rule} is of the form
\[r: H_1 \backslash H_2 \Leftrightarrow C | B\]
where $r:$ is an optional {\em name} (a unique identifier) of a rule.
In the rule {\em head} (left-hand side), $H_1$ and $H_2$ are conjunctions of user-defined constraints,
the optional {\em guard} $C |$ is a conjunction of built-in constraints,
and the {\em body} (right-hand side) $B$ is a goal.

In the rule, $H_1$ are called the {\em kept constraints}, while $H_2$ are called the {\em removed constraints}.
At least one of $H_1$ and $H_2$ must be non-empty. 
If $H_1$ is empty, the rule corresponds to a simplification rule, also written
\[s: H_2 \Leftrightarrow C | B.\]
If $H_2$ is empty, the rule corresponds to a propagation rule, also written
\[p: H_1 \Rightarrow C | B.\]
} 
\end{definition}
Interestingly, most parallel semantics do not allow for propagation rules, while distributed semantics do.
This will be discussed in Section 11.

\myparagraph{Ground CHR} 
Most implementations and some semantics assume that variables are substituted by ground (variable-free) terms at run-time. 
This requirement can be captured by a common syntactic fragment of CHR:
In {\em Ground CHR}, every variable in a rule (also) occurs in the head of the rule.
We also say that the rule is {\em range-restricted}.
This condition can be relaxed by allowing for local variables in the body of rule, provided they first occur in built-in constraints that always bound them to ground values at run-time (e.g. arithmetic functions). 
So given a ground initial states, all states in a computation will stay ground.
As we will see, this greatly simplifies refined semantics and implementations, 
since then it is not necessary to account for the suspension and wake-up of user-defined constraints during computations.
It is worth noting that Ground CHR without propagation rules is still Turing-complete:
it can implement a Turing machine with just one rule 
as we will see in Section \ref{Turing}.

\subsection{Sequential Abstract Operational Semantics of CHR}\label{sec:chr:semantics}

The semantics follows \cite{raiser_betz_fru_equivalence_revisited_chr09,betz2014unified}. 
It relies on a structural equivalence between states that abstracts away from technical details in a transition.

\myparagraph{State Equivalence}
The equivalence relation treats built-in constraints semantically and user-defined constraints syntactically.
Basically, two states are equivalent if they are logically equivalent (imply each other) 
while taking into account that user-defined constraints form a multiset, i.e. multiplicities matter.
For a state $S$, the notation $S_{bi}$ denotes the built-in constraints of $S$
and $S_{ud}$ denotes the user-defined constraints of $S$.
\begin{definition}[State Equivalence]
{\rm
Two states $S_1 = (S_{1bi} \land S_{1ud})$ and $S_2 = (S_{2bi} \land S_{2ud})$ are 
{\em equivalent}, 
written $S_1 \equiv S_2$, if and only if
$$	
\models 
        \forall (S_{1bi} \rightarrow \exists \bar y ((S_{1ud} = S_{2ud}) \land S_{2bi}))
	\land 
         \forall (S_{2bi} \rightarrow \exists \bar x ((S_{1ud} = S_{2ud}) \land S_{1bi}))
$$
\noindent with $\bar x$ those variables that only occur in $S_1$ and $\bar y$ those variables that only occur in $S_2$.
} 
\end{definition}
The CHR state equivalence is defined by two symmetric implications and moreover syntactically equates the conjunctions of user-defined constraints as multisets.
For example, 
$$X{=<}Y \land Y{=<}X \land c(X,Y) \ \equiv \ X{=}Y \land c(X,X) \not\equiv X{=}Y \land c(X,X) \land c(X,X).$$

\myparagraph{Transition}
Using this state equivalence, the abstract CHR semantics is defined by a single transition 
that is the workhorse of CHR program execution. It defines the application of a rule. 
Let the rule $(r : H_1 \backslash H_2 \Leftrightarrow C | B)$ be a variant of a rule from a given program $\cal{P}$.
A variant (renaming) of an expression is obtained by uniformly replacing its variables by fresh variables.

\noindent\rule{\textwidth}{0.5pt}
\begin{center}
{\bf (Apply)} \ \ \ $\underline{S \equiv (H_1 \land H_2 \land C \land G) \ \ \ \ (r : H_1 \backslash H_2 \Leftrightarrow C | B) \in {\mathcal{P}}  \ \ \ \ \ \ \ \ (H_1 \land C \land B \land G) \equiv T}$\\
$S \mapsto_r T$
\rule{\textwidth}{0.5pt}
\end{center}
Upper-case letters stand for (possibly empty) conjunctions of constraints in this section.
The goal $G$ is called {\em context} of the rule application. It is left unchanged. 

In a {\em transition (computation step)}
$S \mapsto_r T$, $S$ is called {\em source state} and $T$ is called {\em target state}.
We may drop the reference to the program $\cal{P}$ and rule $r$ to simplify the presentation.

If the source state can be made equivalent to a state that contains the head constraints and the guard built-in constraints of a variant of a rule, then we delete the removed head constraints from the state and add the rule body constraints to it. Any state that is equivalent to this target state is in the transition relation.

A {\em computation (derivation)} of a goal $S$ in a program $P$
is a connected sequence
$S_i \mapsto S_{i+1}$ beginning with
the {\em initial state (query)} $S_0$ that is $S$ 
and ending in a {\em final state (answer, result)} or the sequence is {\em non-terminating (diverging)}.
The notation $\mapsto^*$ denotes the reflexive and transitive closure of $\mapsto$.

Note that the abstract semantics does not account for termination of {propagation rules}:
If a state can fire a propagation rule once, it can do so again and again, ad
infinitum. This is called trivial non-termination of propagation rules. 
Most parallel semantics rule out propagation rules. 
Propagation rules and their termination will be discussed for distributed CHR in Section \ref{chrd}, though.

For the minimum example, here is a possible {\bf (Apply)} transition from a state $S=(min(0) \land min(2) \land min(1))$ to a state $T=(min(0) \land min(1))$:
\begin{center}
$S \equiv (min(X) \land min(Y) \land X \leq Y \land (X=0 \land Y=2 \land min(1)))$\\ 
$(min(X) \backslash min(Y) \Leftrightarrow X \leq Y | true)$\\ 
$\underline{(min(X) \land X \leq Y \land true \land (X=0 \land Y=2 \land min(1))) \equiv T}$\\
$S \mapsto T$
\end{center}

\subsection{Extension to Parallel Abstract Semantics}

We extend the abstract semantics by parallelism.
We interpret conjunction as parallel operator.
As we have seen for the minimum example, CHR rules can
also be applied simultaneously to overlapping parts of a state, as long as the 
\index{overlap}{\em overlap} (shared, common part)
is not removed by any rule. 
Following \cite{fru_parallel_union_find_iclp05}, CHR parallelism with overlaps is called {\em strong}.
It can be defined as follows, see also Chapter 4 in \cite{fru_chr_book_2009}.

\myparagraph{(Strong) Parallelism (with Overlap)}
We denote parallel transitions by the relation $\parmapsto$.
The transition {\bf (Intro-Par)} says that any sequential transition is also a parallel transition.
The transition {\bf (Parallel)} combines two parallel transitions using conjunction into a single parallel transition 
where the overlap $E$ is kept. 

\noindent\rule{\textwidth}{0.5pt}
\begin{center}
{\bf (Intro-Par)} \ \ \ $\underline{A \mapsto C}$\\
 \ \ \  \ \ \  \ \ \  \ \ \  \ \ \  \ \ \  \ \ \ $A \parmapsto C$\\

\medskip
{\bf (Parallel)} \ \ \ $\underline{A \land E \parmapsto C \land E \ \ \ \ \ \ B \land E \parmapsto D \land E}$\\
 \ \ \  \ \ \  \ \ \  \ \ \  \ \ \  \ \ \  \ \ \ $A \land B \land E \parmapsto C \land D \land E$\\
\rule{\textwidth}{0.5pt}
\end{center}

Again, back to the minimum example:
\begin{center}
(Parallel) \ \ \ $\underline{min(1) \land min(0) \parmapsto true \land min(0) \ \ \ \ \ \ min(2) \land min(0) \parmapsto true \land min(0)}$\\
 \ \ \  \ \ \  \ \ \  \ \ \  \ \ \  \ \ \  \ \ \ $min(1) \land min(2) \land min(0) \parmapsto true \land true \land min(0)$\\
\end{center}
Here the overlap is the goal $min(0)$.

\subsection{Properties: Monotonicity, Soundness and Serializability}

The \index{monotonicity}{\em monotonicity property} of CHR states that adding constraints to a state cannot
inhibit the applicability of a rule \cite{Abdennadher+99}.
It is easy to see from the context of the sequential {\bf (Apply)} transition and from the overlap of the {\bf (Parallel)} transition that a rule can be applied in any state that contains its head and guard.
\begin{theorem} [Monotonicity of CHR] \label{theo:monotonicity}
\index{Monotonicity}
{\rm
   If $A \mapsto B$ then $A \land G \mapsto B \land G$.
   If $A \parmapsto B$ then $A \land E \parmapsto B \land E$.
} 
\end{theorem}

The {\em correctness} of the abstract parallel semantics can be established by proving the following theorem. 
\begin{theorem}[Soundness and Serializability]
{\rm
If $A \parmapsto B$, then there exists a sequential computation $A \mapsto^* B$.
} 
\end{theorem}
The essential aspect of the truth is that the (Parallel) transition can be simulated sequentially:
If    ${{A \land E}}$  $\mapsto$
  ${{B \land E}}$
and  ${{C \land E}}$  $\mapsto$
  ${{D \land E}}$,
then  ${{A} \land {C \land E}}$
  $\mapsto S \mapsto $ 
${{B} \land {D \land E}}$,
where $S$ is either $A\land D \land E$ or $B \land C \land E$, i.e. the two transitions commute.

\section{Parallel CHR Example Programs}

These exemplary CHR programs are mostly folklore in the CHR community, 
see e.g. Chapters 2 and 7 in \cite{fru_chr_book_2009}.
These are concise and effective 
implementations of classical algorithms and problems 
starting with finding primes, sorting,
including Turing machines and
ending with Preflow-Push and Union-Find. 
Often one type of constraint and one rule will suffice, and we will not need more then six rules.
Due to the guaranteed properties of CHR, these programs are also
{incremental anytime online approximation algorithms}. Typically
they run in parallel without any need for modifying the program.
An exception is Union-Find, which is known to be hard to parallelize.
We do it with the help of confluence analysis.

These sequential programs are in the subset of Ground CHR without propagation rules and can therefore be understood in all parallel semantics and 
executed in all parallel implementations surveyed without modification. 
On the other hand, most example programs may require some modification for distributed semantics and their implementations.
As we will see, the experimental results report parallel speed-ups.

\subsection{Algorithms of Erastothenes, Euclid, von Neumann, Floyd and Warshall}

Here we introduce some classical algorithms over numbers and graphs.
They are implemented as simple multiset transformations 
reminiscent of the Chemical Abstract Machine (CHAM).
Typically, they can be implemented with one kind of constraint and a single rule in CHR
that can be applied in parallel to pairs of constraints.
Our running example of minimum falls into this category.
These programs are confluent when run as intended, with ground goals.
Correctness of each implementation can be shown by contradiction: given the specified initial goal, 
if the resulting answer were not of the desired form, the rule would still be applicable.

\myparagraph{Prime Numbers}
The following rule is like the rule for minimum, but the guard is different, more strict.
In effect, it filters out multiples of numbers, similar to the Sieve of Erastothenes.
\begin{verbatim}
sift : prime(I) \ prime(J) <=> J mod I =:= 0 | true.
\end{verbatim}
If all natural numbers from $2$ to $n$ are given, only the prime numbers within this
range remain, since non-prime numbers are multiples of other numbers greater equal to $2$.
Obviously, the rules can be applied to pairs of prime number candidates in parallel.
In a parallel step, we can try to remove each prime by associating it with another prime such that the sift rule is applicable.
This gives a maximum, linear parallel speed-up without the need to modify the program.
This was confirmed experimentally for both a software and a hardware implementation \cite{lam-concurrent-chapter2011,triossi2012compiling}.

\myparagraph{Greatest Common Divisor (GCD)}
The following rule computes the greatest common divisor of natural numbers
written each as {\tt gcd(N)}. 
\begin{verbatim}
gcd(N) \ gcd(M) <=> 0<N,N=<M | gcd(M-N).
\end{verbatim}
The rule replaces $M$ by the smaller number $M-N$ as in Euclid's algorithm.
The rule maintains the invariant that the numbers have the same greatest common divisor.
Eventually, if $N=M$, a zero is produced.
The remaining nonzero {\tt gcd} constraint contains the value of the gcd.
The rules can be applied to pairs of gcd numbers in parallel.
Note that to any pair of gcd constraints, the rule will always be applicable.
A parallel speed-up was observed in a hardware implementation \cite{triossi2012compiling},
and even a super-linear speed-up in a software implementation \cite{lam-concurrent-chapter2011}.

\myparagraph{Merge Sort}
The initial goal state contains arcs of the form {\tt a->V} for each value {\tt V},
where {\tt a} is a given smallest (dummy) value.
\begin{verbatim}
msort : A->B \ A->C <=> A<B, B<C | B->C.
\end{verbatim}
The rule only updates the first argument of the {\tt arc} constraint, never the second.
The first argument is replaced by a larger value and the two resulting arcs form a small chain
{\tt A->B,  B->C}.
The rule maintains the invariant that {\tt A=<B}.
So eventually, in each arc, a number will be followed by its immediate successor, 
and thus the resulting chain of arcs is sorted.

For sorting with optimal run-time complexity, we prefer
merging arc chains of the same length. To this end, we precede each chain with its
length, written as special arc {\tt N=>FirstNode}. We also have to add a rule to initiate merging of
chains of the same length:
\begin{verbatim}
N=>A, N=>B <=> A<B | N+N=>A, A->B.
\end{verbatim}
In the initial goal we now introduce
constraints of the form {\tt 1=>V} for each value {\tt V}.
The rules can be applied to pairs of arcs in parallel similar to the previous examples.

\myparagraph{Floyd-Warshall All-Pair Shortest Paths}
Our implementation finds the shortest distance between all connected pairs of nodes in
the transitive closure of a directed graph whose edges are annotated with non-negative distances.
\begin{verbatim}
shorten : arc(I,K,D1), arc(K,J,D2) \ arc(I,J,D3) <=> 
                            D3>D1+D2 | arc(I,J,D1+D2).
\end{verbatim}
Clearly we can shorten arc distances in parallel by considering triples of arc constraints that match the head of the rule.
In each parallel step, we can try to remove each arc by associating it with a corresponding pair of arc constraints and by checking if the rule is applicable then.

\subsection{Classical Models and Classical Algorithms with Statefulness}

These algorithms about abstract problems are characterized by their {\em statefulness}, i.e. their essence is a state change, an update.
While other declarative languages may not have an efficient way to update, 
CHR has a proven one by constant-time updating (i.e. removing and adding) user-defined constraints
\cite{sney_schr_demoen_chr_complexity_toplas09}.

\myparagraph{Turing Machine}\label{Turing}
The Turing machine is the classical model of computability used in theoretical computer science. One rule suffices to implement it efficiently in CHR.
\begin{verbatim}
st(QI,SI,SJ,D,QJ) \ state(I,QI), cell(I,SI) <=> state(I+D,QJ), cell(I,SJ).
\end{verbatim}
The state transition steps of the Turing machine are given as constraints {\tt st(QI,SI,SJ,D,QJ)}:
in the current state {\tt QI}
reading tape symbol {\tt SI}, write symbol {\tt SJ} and move in direction {\tt D} to be in state {\tt QJ}. 
The direction is either left or right, we move along the cells of a tape.
We represent cells as an array, so positions are numbers and the direction is either $+1$ or $-1$.
A Turing machine with one tape is inherently sequential, since we can only be in one state at a time. 
Still parallelism can be employed to find the matching state transition constraint.

The implementation of the Turing machine shows Turing-completeness of the Ground CHR fragment with constants only and without propagation rules, actually with a single rule \cite{sneyers_subclass_iclp08}.

\myparagraph{Dijkstras Dining Philosophers}
In this classical problem in concurrency, several philosophers sit at a
round table.  Between each of them a fork is placed. A philosopher either
thinks or eats. In order to eat, a philosopher needs two forks, the one from his
left and the one from his right. After a while, an eating philosopher will start
to think again, releasing the forks and thus making them available to his
neighbors again.
\begin{verbatim}
think_eat : think(X), fork(X), fork(Y) <=> Y =:= (X+1) mod n | eat(X).
eat_think : eat(X) <=> Y =:= (X+1) mod n | think(X), fork(X), fork(Y).
\end{verbatim}
In the implementation, we assume a given number {\tt n} of philosophers (and forks).
They are identified by a number from zero to {\tt n-1}.
The rules are inverses of each other, the constraints simply switch sides.

The problem is to design a concurrent algorithm that is fair, i.e. that no philosopher will starve.
Here we are mainly interested in the inherent parallelism of the problem. 
Disjoint pairs of neighboring forks can be used for eating in one parallel computation step.
(For the experiments, time counters for eating and thinking were introduced into the program to introduce termination.)

\myparagraph{Blocks World}
Blocks World is a classical planning problem in Artificial Intelligence. It simulates robot arms re-arranging stacks of blocks.
\begin{verbatim}
grab  : grab(R,X),  empty(R), clear(X), on(X,Y) <=> hold(R,X), clear(Y).
putOn : putOn(R,Y), hold(R,X), clear(Y) <=> empty(R), clear(X), on(X,Y).
\end{verbatim}
The {\em operation constraints} {\tt grab} and {\tt putOn} specify the action that is taken.
The other constraints are {\em data constraints} holding information about the scenario.
Operation constraints update data constraints.
The rule {\tt grab} specifies that robot arm {\tt R} grabs block {\tt X} if 
{\tt R} is empty and block {\tt X} is clear on top and on block {\tt Y}.
As a result, robot arm {\tt R} holds block {\tt X} and block {\tt Y} is clear.
The rule {\tt putOn} specifies the inverse action.
The {data constraints} in the rule switch sides.
At any time, only one of the actions is thus possible for a given robot arm.
Parallelism is induced by introducing several robot arms and multiple actions for them.
Different robot arms can grab different clear blocks in parallel or put different blocks on different clear blocks in parallel.

\subsection{Parallel Preflow-Push Algorithm}

Next we present two non-trivial algorithms, Preflow-Push and Union-Find.
Both algorithms are acknowledged in the literature to be hard to parallelize.
To maintain the focus of the survey, we cannot explain these algorithms in detail.

The Preflow-Push algorithm \cite{GoldbergTarjan} solves the maximum-flow problem. Intuitively the problem can be understood as a system of connected water-pipes, where each pipe has a restricted given capacity. The system is closed except for one source and one sink valve. The problem now is to find the maximum capacity the system can handle from source to sink and to find the routes the water actually takes. 

A flow network is a directed graph, where each edge is assigned a non-negative capacity. 
We want to find a maximum flow through the network from a source to a sink node under the capacity restrictions.
The Preflow-Push algorithm moves flow locally between neighboring nodes until a maximum flow is reached.

In \cite{meister_preflowpush_csclp07}, we present and analyse a concise declarative parallel implementation of the preflow-push algorithm by just four rules.
In the code listing below, comment lines start with the symbol \verb!%!.
\begin{verbatim}
 % increase node height by one, remove minimum
lift : n(U,N), e(U,E) \ h(U,_), m(U,M,C)
       <=> U \= source, U \= sink, 0 < E, C =:= N+E | h(U,M+1).
 % replace K by HU in unchecked egde, insert minimum
up   : h(U,HU), h(V,HV) \ r(U,V,K)
       <=> HU =< HV, K < HU | m(U,HV,1), r(U,V,HU).
 % push flow downwards by one unit, insert minimum, reverse edge
push : h(U,HU), h(V,HV) \ e(U,EU), e(V,EV), r(U,V,_)
       <=>  0 < EU, HV < HU | e(U,EU-1), e(V,EV+1), m(V,HU,1), r(V,U,HV).
 % compute minimum for node, count for completeness
min  : m(U,M1,C1), m(U,M2,C2) <=> m(U,min(M1,M2),C1+C2).
\end{verbatim}
The variable {\tt U} stands for a node, 
{\tt N} is its number of outward capacity edges,
{\tt E} is its current excess flow,
{\tt HU} is its current height.
The constraint {\tt m(U,M,C)} encodes a minimum candidate with value {\tt M} for node {\tt U}, 
where the counter {\tt C} allows to detect if the minimum of all outward edges has been computed.
The constraint {\tt r(U,V,K)} encodes a residual edge from nodes {\tt U} to {\tt V} with remaining capacity {\tt K}.

The {implementation} described in \cite{meister_preflowpush_csclp07} simulates parallel computations sequentially using an {interleaving semantics} approach and time stamps for user-defined constraints. 
The active elements (nodes with excess flow) can be processed in parallel as long as their 
neighborhoods (set of nodes connected to them through an edge) do not overlap.
In the simulation, we greedily, randomly and exhaustively apply as many rules as possible at a given time point $t$ before progressing to time $t+1$. A speed-up in experiments with random graphs was consistently observed. The speed-up depends on the total amount of flow units, its distribution on disjoint nodes, and the density of the flow network. A parallel speed-up was also confirmed in the experiments of \cite{triossi2012compiling}.

\subsection{Parallel Union-Find Algorithm} 

This classical union-find (also: disjoint set union) (UF) algorithm \cite{Tarjan} efficiently
maintains disjoint sets under the operation of union. Each set is represented by a rooted tree, whose
nodes are the elements of the set.
Union-Find is acknowledged in the literature to be hard to parallelize.

In \cite{fru_parallel_union_find_iclp05}, we implement the UF algorithm  in CHR with optimal 
time and space complexity and with the anytime online algorithm properties. 
This effectiveness is believed impossible in other
pure declarative programming languages due to their inability to express \index{destructive
assignment}\index{assignment!destructive}{destructive assignment} in constant time.  
When the UF algorithm is extended by rules that deal with
function terms (rational trees), it can be used for optimal
complexity \index{unification}{\em unification} \cite{meister2007reconstructing}.
Last but not least, 
a generalization of Union-Find yields novel incremental algorithms for simple Boolean and
linear equations \cite{fru_deriving_linear_algorithms_from_uf_chr06}.
See chapter 10 in \cite{fru_chr_book_2009} for an overview of Union-Find in CHR.

\myparagraph{Parallelizing Basic Union-Find}
We only discuss the basic Union-Find (UF) algorithm here, not the optimized one, since the former has been used in experiments \cite{sulz_lam_parallelexecution_ppdp08}.
In CHR, the {\em data constraints} \texttt{root} and arc \texttt{->}
represent the tree data structure.
With the UF algorithm come several {\em operation constraints}:
{\tt find} returns the root of the tree in which a node is contained,
{\tt union} joins the trees of two nodes,
{\tt link} performs the actual join.
\begin{verbatim}
union    : union(A,B) <=> find(A,X), find(B,Y), link(X,Y).

findNode : A->B  \ find(A,X) <=> find(B,X).
findRoot : root(A) \ find(A,X) <=> found(A,X). 

linkEq   : link(X,Y), found(A,X), found(A,Y) <=> true.  
linkRoot : link(X,Y), found(A,X), found(B,Y), root(A) \ root(B) <=> B->A.
\end{verbatim}
The second argument of the find operation \texttt{find} holds a fresh variable as identifier.
When the root is found, it is recorded in the constraint \texttt{found}.

CHR {confluence analysis} \cite{abd_fru_integration_lopstr03,abd_fru_completion_cp98}
produces abstract states that reveal a {deadlock}:
when we are about to apply the {\tt linkRoot} rule,
another link operation may remove one of the roots that we need for
linking. 
From the non-confluent states we can derive an additional rule for {\tt found} that mimics the rule 
{\tt findNode}: the {\tt found}
constraint now keeps track of the updates of the tree so that its result argument is
always a root.
\begin{verbatim}
foundUpdate : A->B \ found(A,X) <=> found(B,X).
\end{verbatim}
Linking for disjoint node pairs can now run in parallel.  
While this seems an obvious result, this semi-automatic confluence-based approach 
yields a non-trivial parallel variant of the optimized UF algorithm with path compression.
{Correctness} of the parallelisation is proven in both cases in \cite{fru_parallel_union_find_iclp05}.
A parallel speed-up is reported in \cite{lam-concurrent-chapter2011}.

\section{Parallel CHR with Transactions}\label{transactionCHR}

We now extend parallel CHR by transactions.
Transactions will also be used for the implementation of parallel CHR in Section \ref{sec:impt-par-chr-ghc}
and for encoding of a transaction-based
concurrency model in CHR in Section \ref{subsec:stmchr}.

\myparagraph{Transactions} 
They alleviate the complexity of writing concurrent programs by offering entire computations to run 
atomically and in isolation. 
{\em Atomicity} means that a transaction either proceeds un-interrupted and successfully commits or has to rollback (undo its side-effects).
In {\em optimistic concurrency control}, updates are logged and only committed at the end of a transaction when there are no update conflicts with other transactions.
{\em Isolation} means that no intermediate update is observable by another transaction.
The highest level of isolation is {\em serializability}, the major correctness criterion for concurrent transactions: for each parallel execution there is a sequential execution with the same result.

\subsection{Transactions in Parallel CHR}

The paper \cite{schr_sulz_transactions_iclp08} 
proposes \chrt\ as a conservative extension of CHR with atomic transactions. An atomic transaction is denoted as a meta-constraint {\tt atomic(C)} where {\tt C} is a conjunction of CHR constraints. Atomic transactions may appear in goals.

\begin{example}
{\rm
Consider these CHR rules for updating a bank account:
\begin{verbatim}
balance(Acc,Bal), deposit(Acc,Amt)  <=> balance(Acc,Bal+Amt).
balance(Acc,Bal), withdraw(Acc,Amt) <=> Bal>Amt | balance(Acc,Bal-Amt).

transfer(Acc1,Acc2,Amt) <=> withdraw(Acc1,Amt), deposit(Acc2,Amt).
\end{verbatim}
The {\tt balance} constraint is a {\em data constraint}, and the {\tt deposit} and {\tt withdraw} constraints are {\em operation constraints}. 
The guard ensures that withdrawal is only possible if the amount in the account is sufficient. 
The transfer constraint rule combines deposit and withdrawal among two accounts.

Now assume a transfer between two accounts: 
\begin{verbatim}
balance(acc1,500), balance(acc2,0), transfer(acc1,acc2,1000)
\end{verbatim}
We can execute the deposit, but we cannot execute the withdrawal due to insufficient funds. 
The transaction gets {\em stuck}. It has a {deadlock} and cannot proceed till the end. 
This is clearly not the desired behavior of a transfer.

In \chrt, we can introduce a transaction to avoid this problem.
The {\tt transfer} constraint in the goal is wrapped by the meta-constraint {\tt atomic}.
\begin{verbatim}
balance(acc1,500), balance(acc2,0), atomic(transfer(acc1,acc2,1000)) 
\end{verbatim}
Now the incomplete transaction will be rolled back, no money will be transferred.
} 
\end{example}

\subsection{Abstract Syntax and Semantics of \chrt}

We assume {Ground CHR}.
We classify CHR constraints into {\em operation constraints} and {\em data constraints}.
The distinction appeals to the intuitive understanding that
operation constraints update data constraints.
Thus the head of a \chrt\ rule must contain exactly one operation constraint.
It requires one more transition for transactions.
The {\bf (Atomic)} transition executes any number of atomic transactions in parallel in a common context $T$ of data constraints.

\noindent\rule{\textwidth}{0.5pt}
\begin{center}
{\bf (Atomic)}   \ \  \ \ \  \ \  $\underline{  \ \  \ \ \  \ \ \ \ \ \ \ (T \land S_1 \land C_1 \mapsto^* T \land S'_1), \ldots (T \land S_n \land C_n \mapsto^* T \land S'_n)}$\\
  \ \  \ \ \  \ \ \ \  \ \  \ \ \  \ \  \  \ \ \   $T \land S_1 \land \ldots S_n \land atomic(C_1) \land \ldots atomic(C_n) \parmapsto T \land S'_1 \land \ldots S'_n$\\
\rule{\textwidth}{0.5pt}
\end{center}
In the transition, $T, S_i,$ and $S'_i$ must be data constraints. 
The parallel step considers the separate evaluation of each $C_i$ in isolation. 
The transactions only share the common data constraints $T$, which serves as a context. 
Note that each transaction may perform arbitrary many computation steps. 
Each transaction is fully executed until there are no operation constraints.
It does not get {stuck}.
So there are only data constraints in the target state.

\subsection{Properties: Monotonicity, Soundness and Serializability}

For \chrt\ programs, the following properties are proven to hold in \cite{schr_sulz_transactions_iclp08}.
\begin{description}
\item[Serializability]
For each (Atomic) transition with $n$ concurrent transactions, there is a corresponding computation of $n$ consecutive sequential (Atomic) transitions each with only one transaction.

\item[Soundness]
For any computation in \chrt, there is a corresponding computation in CHR where the atomic wrappers are dropped.

\item[Monotonicity]
Although not proven in the paper, it follows from Soundness and the context $T$ of the (Atomic) transition.
\end{description}

\subsection{Encoding Transactions in Standard CHR}

We want to execute \chrt\ in standard parallel CHR, i.e without the (Atomic) transition. 
The straightforward way is to execute atomic transactions only sequentially. 
Thus, we trivially guarantee the atomic and isolated execution of transactions.
We identify two special cases where we can erase the atomic wrappers and still allow for parallel execution: 
bounded and for {confluent} transactions.

\myparagraph{Bounded Transactions}
A {\em bounded transaction} is one that performs a finite, statically known number of transitions.
We eliminate a bounded transaction {\tt atomic(G)} from a program by adding a rule to the program of the form 
{\tt atomic(G) <=> G}.
Then we {unfold the rule} \cite{fru_specialization_lopstr04,gabbrielli_unfolding_tplp13,fru_holz_source2source_agp03} until no more operation constraints appear in its body.
Since the transaction is bounded, unfolding will eventually stop.

In the running example, we can replace the atomic transfer rule (since it is bounded) by the following rule.
\begin{verbatim}
balance(Acc1,Amt1),balance(Acc2,Amt2),atomic(transfer(Acc1,Acc2,Amt)) <=> 
             Amt1>Amt | balance(Acc1,Amt1-Amt), balance(Acc2,Amt2+Amt).
\end{verbatim}
The rule head expresses the fact that an atomic transfer requires exclusive access to both accounts involved.

\myparagraph{Confluent Transactions}
The paper proves that if a \chrt\ program is confluent when we ignore atomic wrappers,
then it can be executed in standard parallel CHR provided the initial goal never gets stuck (deadlocks).
{Confluence} then guarantees that isolation is not violated.

Consider the example of the {stuck} transaction that attempts to overdraw an account. 
The withdraw rule can be fixed if we drop its guard (and hence allow negative balances):
\begin{verbatim}
balance(Acc,Bal), withdraw(Acc,Amt) <=> balance(Acc,Bal-Amt).
\end{verbatim}
Any two consecutive transfers commute now. Regardless of the order they are performed in, they yield the same final result
(even if the intermediate results differ). Hence, we can safely erase the atomic wrappers.

\section{Refined Parallel CHR Semantics} \label{sec:lam-par-ref}

A {\em refined} semantics for parallel CHR is developed and implemented in
\cite{
sulz_lam_parallelexecution_ppdp08,lam_sulzmann_conc_goal_based_tplp09,lam-concurrent-chapter2011}. 
This semantics can be seen as a refinement of the parallel abstract semantics given before. 
In states, we now differentiate between the goal that holds active constraints to be processed, and the constraint store that holds inactive suspended constraints as data.
This means that we have to account for the in-activation (suspension) and re-activation (wake-up) of user-defined constraints due to built-in constraints on shared variables.
As before, the semantics is given in two parts, the sequential transitions and the parallel transitions 
and the properties of monotonicity, soundness and serializability are shown.

\subsection{Syntax for Refined Parallel CHR} 

\ELfig{fig:chr-goal-syntax}{Refined Parallel CHR Syntax}{
{
\bda{l}
 \ba{lllll}
  \mbox{Built-In Constraint}       & e \\ 
  \mbox{CHR Constraint} & c ::= p(\overline{t}) & &
  \mbox{Identified (CHR) Constraint} & nc ::= c\#i         \\ 
  \mbox{Goal Constraint} & g ::= c \mid e \mid nc & &
  \mbox{Goal (Store)} & G ::= \biguplus {g}       \\ 
  \mbox{Store Constraint} & sc ::= e \mid nc & &  
  \mbox{(Constraint) Store} & Sn ::= \bigcup {sc} \\
  \mbox{State} & \sigma ::= \langle G,Sn \rangle & & 
  \mbox{Matched Constraints} & \delta ::= Sn ~\backslash~ Sn
  \ea
\eda
}
}

Figure \ref{fig:chr-goal-syntax} describes the syntax for the refined semantics. 
The notation $\overline{a}$ denotes a sequence of $a$'s.
We only consider built-in constraints that are syntactic equalities or arithmetic equations.
To distinguish multiple occurrences (copies, duplicates) of CHR constraints,
they are extended by a unique identifier. We call
$c\#i$ an {\em identified constraint}.
Conjunctions are modeled as (multi)sets.
Unlike in the abstract semantics, a state is now a pair:
we distinguish between a goal (store) (a multiset of constraints) and the (constraint) store (a set of built-in and identified CHR constraints).
Correspondingly, there are goal and store constraints.
We also introduce matched constraints that are pairs of store constraints which we will need as an annotation to transitions.

\subsection{Sequential Refined CHR Semantics} \label{sec:conc-goal-based-chr-sem}

The sequential part of the semantics in Figure~\ref{fig:goal-semantics} 
is a generalization of the refined CHR semantics of~\cite{duck_stuck_garc_holz_refined_op_sem_iclp04}.
The semantics assumes generalized simpagation rules that are not {propagation rules}.

Constraints from the goal are executed one by one.
A constraint currently under execution is called {\em active constraint}.
It tries to apply rules to itself.
To try a rule, the active constraint is matched against a head
constraint of the rule.  The remaining head constraints are matched with 
{\em partner constraints} from the constraint store.
If there is such a complete matching and
if the guard is satisfied under this matching, then the rule applies
(fires). The constraints matching the removed constraints of the head 
are deleted atomically and the body of the rule is added to the state.
Because of the role of the active constraint, we call the semantics {\em goal-based semantics}.

\ELfig{fig:goal-semantics}{Parallel CHR Semantics (Sequential Part $\ELgoaltranssf{\delta}$)}{
{
\index{State Transition System!Goal-Based}
\bda{c} 
   \ba{ccc}
      \ELtlabel{\bf Solve+Wake} & & 
      \ELmyirule{W = WakeUp(e,Sn)}
      {\ELchrstate{\{e\} \uplus G}{Sn} \ELgoaltranssf{W \backslash \{\}} \ELchrstate{W \uplus G}{\{e\} \ELstcup Sn}} \\ \\
      \ELtlabel{\bf Activate} & &
      \ELmyirule{i \mbox{ is a fresh identifier}}
              {\ELchrstate{\{c\} \uplus G}{Sn} \ELgoaltranssf{\{\} \backslash \{\}} \ELchrstate{\{c\#i\} \uplus G}{\{c\#i\} \ELstcup Sn}} \\ \\
      \ELtlabel{\bf Apply-Remove} & &
      \ELmyirule{ \mbox{Variant of } (r~\atsign~H_P' \backslash H_S' \ELsimparrow t \mid B') \in {\cal P} \mbox{ such that} \\
                \exists \phi \ELsgap Eqs(Sn) \models \phi(t) \ELsgap \phi(H_P') = DropIds(H_P) \\
                        \phi(H_S') = \{c\} \uplus DropIds(H_S) \ELsgap
                \delta = H_P \backslash \{c\#i\} \ELstcup H_S }
              {\ba{ll}
                & \ELchrstate{\{c\#i\} \uplus G}{\{c\#i\} \ELstcup H_P \ELstcup H_S \ELstcup Sn} \\
               \ELgoaltranssf{\delta} & \ELchrstate{\phi(B') \uplus G}{H_P \ELstcup Sn}
               \ea } \\ \\
      \ELtlabel{\bf Apply-Keep} & &
      \ELmyirule{ \mbox{Variant of } (r~\atsign~H_P' \backslash H_S' \ELsimparrow t \mid B') \in {\cal P} \mbox{ such that} \\
                \exists \phi \ELsgap Eqs(Sn) \models \phi(t) \ELsgap \phi(H_S') = DropIds(H_S) \\
                        \phi(H_P') = \{c\} \uplus DropIds(H_P) \ELsgap
                \delta = \{c\#i\} \ELstcup H_P \backslash H_S }
              {\ba{ll}
                & \ELchrstate{\{c\#i\} \uplus G}{\{c\#i\} \ELstcup H_P \ELstcup H_S \ELstcup Sn} \\
               \ELgoaltranssf{\delta} & \ELchrstate{\phi(B') \uplus \{c\#i\} \uplus G} 
                                               {\{c\#i\} \ELstcup H_P \ELstcup Sn}
               \ea } \\ \\
      \ELtlabel{\bf Suspend} & &
      \ELmyirule{\ELtlabel{Apply-Remove} \mbox{ and } \ELtlabel{Apply-Keep} \mbox{ do not apply to } c\#i \mbox{ in } Sn}
              {\ELchrstate{\{c\#i\} \uplus G}{Sn} \ELgoaltranssf{\{\} \backslash \{\}} 
               \ELchrstate{G}{Sn}}
   \ea
 \\ \\   
   \ba{clll}
   \mbox{where} 
         & Eqs(S)       & = & \{e \mid e \in Sn, e \mbox{ is a built-in costraint}\} \\       
         & DropIds(Sn)  & = & \{ c \mid c\#i \in Sn \} \uplus \{ e \mid e \in Sn\} \\
         & WakeUp(e,Sn) & = & \{ c\#i \mid c\#i \in Sn \wedge \phi \mbox{ m.g.u. of } Eqs(Sn) \wedge \\
         &              &   & \theta \mbox{ m.g.u. of } Eqs(Sn \cup \{e\}) \wedge \phi(c) \neq \theta(c) \} \\
   \ea
\eda
} 
}

\myparagraph{Transitions}
A transition
$\sigma \ELgoaltranssf{\delta} \sigma'$ maps the CHR state $\sigma$ to $\sigma'$ 
involving the CHR constraint goals in $\delta$. 
The transition annotation $\delta$ holds the constraints that where matched with the rule head. 
It will be needed in the parallel part of the semantics.

The first transition \ELtlabel{Solve+Wake} moves a built-in constraint, an equation or equality $e$, 
into the store and wakes up (reactivates)
identified constraints in the store which could now participate in a rule application. 
This is the case when the built-in constraint effects variables in a user-defined constraint, 
because then the re-activated (woken) constraint may now be able to match a rule head and satisfy the guard of the rule.
The function $WakeUp(e,Sn)$ computes a conservative approximation
of the reactivated constraints,
where m.g.u. denotes the most general unifier induced by a set of syntactic equations.

In transition \ELtlabel{Activate}, 
a CHR constraint goal becomes active by annotating it with a fresh unique identifier and adding it 
to the store. 

Rules are applied in transitions \ELtlabel{Apply-Remove} and \ELtlabel{Apply-Keep}. 
They are analogous, but distinguish if the active constraint $c\#i$ is kept or removed. 
In both cases, 
we seek for the missing {partner constraints} in the store, producing a {\em matching substitution} $\phi$ in case of success. 
The guard $t$ must be logically entailed by the built-in constraints in the store under the 
substitution $\phi$.
Then 
we apply the rule instance of $r$ by atomically removing
the matching constraints $H_S$ and adding the rule body instance $\phi(B)$ to the goal.
We also record the matched identified constraints $H_S$ and $H_P$ in the transition annotation. 
In transition \ELtlabel{Apply-Remove}, 
the matching constraints $H_S$ include $c\#i$. Since 
$c\#i$ is removed, we drop it from both the goal and the store. 
In transition \ELtlabel{Apply-Keep}, $c\#i$ remains and so 
can possibly fire further rules. 

Finally, in transition \ELtlabel{Suspend}, we put an active constraint to sleep. 
We remove the active identified constraint from the goal
if no (more) rules apply to the constraint. 
Note that the constraint is kept suspended in the store and may be woken later on.

\subsection{Extension to Parallel Refined CHR Semantics} 

Figure \ref{fig:conc-goal-semantics} presents the  parallel part of the refined operational 
semantics. It is a refinement of the parallel transition for the abstract semantics.
We allow for multiple goal stores to be combined while the constraint store is shared among the parallel computations.

\ELfig{fig:conc-goal-semantics}{Parallel CHR Semantics (Parallel Part $\ELpartranssf{\delta}$)}{
 \bda{c} \index{State Transition System!Goal-Based}
   \ba{ccc}
      \ELtlabel{\bf Intro-Par} & &
      \ELmyirule{\ELchrstate{G}{Sn} \ELgoaltranssf{\delta} \ELchrstate{G'}{Sn'}}
              {\ELchrstate{G}{Sn} \ELpartranssf{\delta} \ELchrstate{G'}{Sn'}} \\ \\
      \ELtlabel{\bf Parallel-Goal} & & 
      \ELmyirule{\ELchrstate{G_1}{H_{S1} \ELstcup H_{S2} \ELstcup Sn} \ELpartranssf{\delta_1}
               \ELchrstate{G_1'}{H_{S2} \ELstcup Sn} \\
               \ELchrstate{G_2}{H_{S1} \ELstcup H_{S2} \ELstcup Sn} \ELpartranssf{\delta_2} 
               \ELchrstate{G_2'}{H_{S1} \ELstcup Sn} \\
               \delta_1 = H_{P1} \backslash H_{S1} \ELsgap \delta_2 = H_{P2} \backslash H_{S2} \\
               H_{P1} \subseteq Sn \ELsgap H_{P2} \subseteq Sn \ELsgap \delta = H_{P1} \cup H_{P2} \backslash H_{S1} \cup H_{S2} \\
               H_{S1} \cap (H_{P2} \cup H_{S2}) = \{\} \ELsgap  H_{S2} \cap (H_{P1} \cup H_{S1}) = \{\} }
              {\ba{ll}
                 & \ELchrstate{G_1 \uplus G_2 \uplus G}{H_{S1} \ELstcup H_{S2} \ELstcup Sn} 
                \ELpartranssf{\delta} 
                     \ELchrstate{G_1' \uplus G_2' \uplus G}{Sn}
               \ea} 
   \ea
 \eda
}

In the \ELtlabel{Intro-Par} transition, we turn a sequential computation into a parallel computation. 
Transition \ELtlabel{Parallel-Goal} parallelizes two parallel computations operating on the same shared store, if
their matched constraints $\delta_1$ and $\delta_2$ do not have an overlap that involves removed constraints.
They may overlap in the kept constraints.
This makes sure that parallel computations remove distinct constraints in the store.
The identifiers of constraints make sure that we can remove multiple but different copies of the same constraint.
The matched constraints $\delta_1$ 
and $\delta_2$ are composed by the union of the kept and removed components, respectively, 
forming $\delta$.
Note that a context $G$ is added to the goals in the resulting parallel transition, implying monotonicity.

\subsection{Properties: Monotonicity, Soundness and Serializability} \label{ssec:corr}

The following results are proven in the appendix of \cite{lam_sulzmann_conc_goal_based_tplp09}.

{\em Monotonicity} holds for the goal store, but not for the constraint store.
In an enlarged constraint store, the \ELtlabel{Suspend} transition may not be possible anymore, because a new rule becomes applicable to the active constraint. 
The monotonicity is still sufficient though, because in the semantics, 
the constraint store is only populated via the goal store.
{\em Serializability} holds:
Any parallel computation can be simulated by a sequence of sequential computations in the refined semantics.

Furthermore, {\em soundness} holds:
any parallel computation has a correspondence in a suitable variant of the sequential abstract semantics.
For the upcoming theorem, let us note that
an {\em initial state} is of the form $\langle G,\{\} \rangle$, 
a {\em final state} is of the form $\langle \{\},Sn \rangle$.
Given a computation $\ELchrstate{G}{\{\}} \ELpartransstar \ELchrstate{G'}{Sn}$, the state $\ELchrstate{G'}{Sn}$ 
is called a {\em reachable state}.
\begin{theorem}[Soundness]
{\rm
   For any reachable state $\ELchrstate{G}{Sn}$, 
   \bda{ll} 
    \mbox{if }   & \ELchrstate{G}{Sn} \ELpartransstar \ELchrstate{G'}{Sn'} \\
    \mbox{then } 
                 & (NoIds(G) \uplus DropIds(Sn)) \mapsto^* (NoIds(G') \uplus DropIds(Sn'))
   \eda
   where $\mapsto^*$ denotes transitions in the sequential abstract semantics
and where $NoIds = \{c \mid c \in G, c \mbox{ is a CHR constraint}\} 
          \uplus \{e \mid e \in G, e \mbox{ is a built-in constraint}\}$.
} 
\end{theorem}

\section{Parallel CHR Implementation in Haskell} \label{sec:impt-par-chr-ghc}

The parallel refined semantics from the previous Section \ref{sec:lam-par-ref}
has been implemented in the lazy functional programming language Haskell
\cite{sulz_lam_lazy_concurr_search_chr07,lam_sulz_concurrent_chr_damp07,sulz_lam_parallelexecution_ppdp08,lam_sulzmann_conc_goal_based_tplp09,lam-concurrent-chapter2011}.
Concretely, we use the Glasgow Haskell Compiler for implementing parallel Ground CHR
because of its good support for shared memory and multi-core architectures. 
The implementation is available online for free download at
\url{https://code.google.com/archive/p/parallel-chr/}.
In principle, the system can be reimplemented in mainstream procedural languages such as C and Java. 
In this section we give an overview of the implementation principles and the best experimental results, details and more experiments with different settings can be found in the literature cited above.

\subsection{Implementation Principles} \label{ssec:conc-prim-choice}

Our implementation follows the principles of standard sequential implementations of CHR where possible
\cite{holz_garc_stuck_duck_opt_comp_chr_hal_tplp05,vanweert_lazy_evaluation_tkde10}.
The goal store is realized as a stack, the constraint store as a hash table. 
We implement common CHR optimizations, such as
constraint indexing (hashing) and 
optimal join ordering for finding partner constraints with early guard scheduling.

Parallel goal execution must not remove constraints in overlaps that participate in several rule head matchings.
We discuss two approaches of concurrency control to implement this kind of parallel rule-head matching, 
locking and transactions, before we settle for a hybrid approach.

\myparagraph{Fine-grained Lock-based Parallel Matching}
Pessimistic concurrency control uses locking as the basic serialization mechanism.
We restrict the access to each constraint in the shared store with a lock. 
When an {active} constraints finds an applicable rule,
it will first try to lock its matching removed {partner} constraints. 
Kept constraints can be used by several rules simultaneously, so they need not be locked. 
Locking fails if any constraint in the complete rule head matching is already locked by another active constraint. 
If locking fails, the active constraint releases all its locks and tries to redo the rule application.
If locking succeeds, the rule is applied.
No unlocking is necessary since locked constraints are removed.
This locking mechanism can avoid deadlocks and cyclic behavior using standard
techniques for these problems such as timestamps or priorities.

\myparagraph{Software Transactional Memory (STM)}
Optimistic concurrency control is based on transactions that can either commit or rollback and restart.
We use the STM transactions provided in Haskell.
The principles of transaction have been introduced in Section \ref{transactionCHR}.
The idea of STM is that atomic program regions are executed optimistically. 
That is, any read/write operations performed by the region are recorded 
locally and will only be made visible when the transaction is completed. Before making the changes 
visible, the underlying STM protocol will check for read/write conflicts with other atomically 
executed regions. 
If there are update conflicts among transactions, the STM protocol will randomly
commit one of the atomic transactions and rollback the others \cite{Shavit1997}. 
Committing means that the programs updates become globally visible. 
Rollback means that we restart the program.  
The disadvantage of STM is that unnecessary rollbacks can happen.
We will meet STM again in Section \ref{subsec:stmchr}, when it is specified in CHR.

\myparagraph{Hybrid STM-based Locking Scheme}
In the implementation, we use both Software 
Transactional Memory (STM) and traditional shared memory access locking techniques. The search for matching 
partner constraints is performed outside STM to avoid unnecessary rollbacks. 
When a complete rule head matching is found, we perform an STM 
procedure that we call {\em atomic rule-head verification (ARV)}.
It checks that all the constraints are still available and marks the constraints to be removed as deleted.  
These deleted constraints will be physically 
delinked from the constraint store, either immediately or later. Both behaviors can be implemented with standard 
concurrency primitives (such as compare-and-swap and locks).

\myparagraph{Thread Pool}
The naive way to implement a parallel CHR system is to {spawn an active thread} for each goal constraint in a state. Each thread tries to find its {partner constraints}. However, the thread and its later partner constraints would then compete for the same rule application. Moreover, the number of threads would be unbounded, as the number of constraints in a state is unbounded.
Our implementation uses a bounded number of active threads. 
A {\em thread pool} maintains threads waiting for tasks to be allocated for  parallel execution.

\subsection{Experimental Results} \label{ssec:impl-parachr-results}

Experiments were performed on an Intel Core quad core processor 
with hyper-threading technology (that effectively allows it to run 8  parallel threads) .
We measure the relative
performance of executing with $1$, $2$, $4$, $8$ and an unbounded number of threads against our
sequential CHR implementation in Haskell. 
The table in Figure \ref{fig:optimal-results} gives some exemplary results with 
these two optimizations:
each goal thread searches store constraints in a unique order to avoid matching conflicts
and a special goal ordering for Merge Sort and Gcd is used (explained below).

\begin{figure}[htb]
\begin{center}
{
\begin{tabular}{| l || c | c | c | c | r |}
  \hline 
   Number of Threads   & 1        & 2         & 4         & 8         & Unbounded \\ \hline 
   Merge Sort          & 121\%    & 94\%      & 70\%      & 52\%      & $>$200\% \\ 
   Gcd                 & 109\%    & 37\%      & 18\%      & 12\%      & 123\%     \\ 
   Parallel Union-Find & 125\%    & 82\%      & 52\%      & 32\%      & $>$200\% \\ 
   Blocks World          & 123\%    & 77\%      & 54\%      & 39\%      & $>$200\% \\ 
   Dining Philosophers & 119\%    & 74\%      & 49\%      & 41\%      & $>$200\% \\ 
   Prime               & 115\%    & 73\%      & 46\%      & 30\%      & 155\%     \\ 
   Fibonacci           & 125\%    & 85\%      & 59\%      & 39\%      & $>$200\% \\ 
   Turing Machine      & 111\%    & 63\%      & 78\%      & 70\%      & $>$200\% \\ \hline
\end{tabular}
}
\end{center}
\caption{Experimental results, with optimal configuration (on 8 threaded Intel processor)}
\label{fig:optimal-results}
\end{figure}

There are several general observations to be made with regard to the number of threads. 
Executing with one
goal thread is clearly inferior to the sequential implementation because of the wasted
overhead of parallel execution. 
Executions with $2$, $4$ and $8$ goal threads show a consistent parallel speed-up, with exception of the Turing Machine. It 
is inherently single-threaded. Interestingly, we still obtain improvements from parallel execution 
of administrative procedures (for example dropping of goals due to failed matching). 
Unbounded thread pooling is always slower than the sequential implementation.
Furthermore we observed a super-linear speed-up for the Gcd example 
with a {\em queue-based goal ordering} instead of the usual {\em stack-based ordering} in the goal store.
In merge sort, we stack {\tt ->} constraints and queue just {\tt =>} for optimal performance.
Last but not least, experiments also confirmed that there is a speed-up when a multi-core processor instead of a single-core processor is used.

\section{Massively-Parallel Set-Based CHR Semantics}

\newcommand{\prim}{prime}
\newcommand{\scc}{scc}
\newcommand{\del}{del}

\newcommand{\stmod}[3]{\ensuremath{\tuple{#1; #2}}}
\newcommand{\stmp}[3]{\ensuremath{\tuple{#1; #2}}}
\newcommand{\stmps}[3]{\ensuremath{\tuple{\{#1\}; #2}}} 

\newcommand{\dermp}{\ensuremath{\twoheadrightarrow}}

\newcommand{\mcD}{\ensuremath{\mathcal{D}}}
\newcommand{\mcR}{\ensuremath{\mathcal{R}}}
\newcommand{\mcO}{\ensuremath{\mathcal{O}}}

A CHR semantics is {\em set-based} if conjunctions of constraints are considered as set instead of multiset.
In \cite{raiser_fru_parallel_wlp10}, we present a parallel execution strategy for set-based CHR. The use of sets instead of multisets has a dramatic impact: it allows for multiple removals of constraints. This means that overlaps can be removed several times. We show that the resulting refined semantics is not sound in general anymore, but sound if the program is deletion-acyclic (i.e., when its simpagation rules do not allow for mutual removal of constraints).
\chrmp\ programs for the computation of minimum, prime numbers, and sorting can run in constant time, given enough processors.
We describe a program for SAT solving in linear time.

\subsection{Massively-Parallel Set-Based Semantics \chrmp}
\label{sec:parallel}

As in the parallel abstract semantics, there are no restrictions on the syntax of CHR.
Reconsider the essential (Parallel) transition of the abstract CHR semantics.
Keep in mind that conjunctions of constraints are now interpreted as sets of constraints.

\begin{center}
(Parallel) \ \ \ $\underline{A \land E \parmapsto C \land E \ \ \ \ \ \ B \land E \parmapsto D \land E}$\\
 \ \ \  \ \ \  \ \ \  \ \ \  \ \ \  \ \ \  \ \ \ $A \land B \land E \parmapsto C \land D \land E$\\
\end{center}
Consider the program
\begin{verbatim}
                      a <=> b,c.           a <=> b,d.
\end{verbatim}
Then the following transition for the goal {\tt a} $\land$ {\tt e} is possible in the set-based interpretation:
\begin{center}
 \ \ \  \ \ \  \ \ \  \ \ \ $\underline{{\tt a} \land {\tt e} \parmapsto {\tt b} \land {\tt c} \land {\tt e} \ \ \ \ \ \ {\tt a} \land {\tt e} \parmapsto {\tt b} \land {\tt d} \land {\tt e}}$\\
 \ \ \  \ \ \  \ \ \  \ \ \  \ \ \  \ \ \  \ \ \ ${\tt a}  \land {\tt e} \parmapsto {\tt b} \land {\tt c} \land {\tt d} \land {\tt e}$\\
\end{center}
This means that {\tt a} is removed twice and {\tt b} is only produced once.

When we generalize this observation, we see that overlaps between rule matchings can be removed arbitrary many times,
leading to a kind of massive parallelism.

\myparagraph{Refined \chrmp\ Semantics}
We refine this set-based semantics now.
We assume CHR without {propagation rules}. 
In the body of a rule, we distinguish between CHR constraints $B_c$ and built-in constraints $B_b$, and write $B_c, B_b$.
A \chrmp\ state $S$ (or $T$) is of the form $\stmod{\bbG}{\bbB}{\bbV}$, where the goal (store) $\bbG$ is a {\em set} (not multiset) of constraints
and the (built-in) constraint store $\bbB$ is a conjunction of built-in constraints.
$c$ and $d$ are atomic constraints.
We adapt the state equivalence $\equiv$ in the obvious way to \chrmp\ states.

\begin{definition}[Massively Parallel Transition] \label{def:dermp}
{\rm
Given a \chrmp\ state $S = \stmod{\bbG}{\bbB}{\bbV}$.
Let $\mcR$ be the
smallest set such that for each rule variant~$r : H_1 \backslash H_2 \Leftrightarrow G \mid B_c, B_b$ 
where $S \equiv \stmod{H_1 \cup H_2 \cup \bbG'}{G \land \bbB'}{\bbV}$
it holds that $(H_1,H_2,B_c,B_b,\bbB') \in \mcR$. 
We then define for any non-empty subset $R \subseteq \mcR$:
\begin{description}
  \item[-] the set of removed constraints $D = \{ c \mid \exists
  (\_,H_2,\_,\_,\bbB') \in R, c \in \bbG : H_2 \land \bbB' \rightarrow c \}$
  \item[-] the set of added constraints $A = \{ c \mid \exists (\_,\_,B_c,\_,\_)
  \in R : c \in B_c \}$
  \item[-] the conjunction of added built-in constraints $B =
  \bigwedge\limits_{(\_,\_,\_,B_b,\bbB') \in R} \bbB' \land B_b$
\end{description}
A \emph{massively parallel transition (step)} of $S = \stmod{\bbG}{\bbB}{\bbV}$ using $\mcR$ is then defined as:

\noindent\rule{\textwidth}{0.5pt}
\[
\mbox{ (Massive-Apply) } \ \ \  \stmod{\bbG}{\bbB}{\bbV} \dermp^R \stmod{(\bbG \setminus D) \cup A}{\bbB \land B}{\bbV}
\] 
\rule{\textwidth}{0.5pt}
If the specific set~$R$ is not of importance we write $\dermp$ instead of $\dermp^R$.
} 
\end{definition}

The idea is that in the set $\mcR$ we collect all possible rule applications and 
then we apply any subset of them at once in one parallel computation step.
In this way, multiple removals of the same constraint are possible.
In the extreme case, $R = \mcR$, so all possible rule applications are performed simultaneously.
We call this {\em exhaustive parallelism}.
With such an execution strategy, any \chrmp\ program is trivially {\em confluent}, 
because there are no conflicting rule applications.
On the other hand, if $R$ is a singleton set, only one rule is applied and we are back to sequential CHR.

\begin{example}
{\rm
Reconsider the CHR program for computing prime numbers.
Consider the state
$$S = \stmp{\{\prim(2),\prim(3),\prim(4),\prim(5),\prim(X)\}}{X{=}6}{\{X\}}.$$
There are three possible rule applications, removing the non-prime
numbers $4$ and twice $6$: 
\[
\mcR = \left\{
\begin{array}{l}
(\{\prim(N_1)\},\{\prim(M_1)\},\emptyset,\top,X{=}6 \land N_1{=}2 \land M_1{=}4),\\
(\{\prim(N_2)\},\{\prim(M_2)\},\emptyset,\top,X{=}6 \land N_2{=}2 \land M_2{=}6),\\
(\{\prim(N_3)\},\{\prim(M_3)\},\emptyset,\top,X{=}6 \land N_3{=}3 \land M_3{=}6)
\end{array}
\right\}
\]
We can now perform all three possible rule applications exhaustively parallel, i.e. $R =
\mcR$, resulting in the following sets:
\[
\begin{array}{l} 
D = \{\prim(4),\prim(X)\}, \ \ 
A = \emptyset,\\
B = (X{=}6 \land N_1{=}2 \land M_1{=}4) \land (X{=}6 \land N_2{=}2 \land
M_2{=}6) \land (X{=}6 \land N_3{=}3 \land M_3{=}6)
\end{array}
\]
This leads to the parallel transition:
\[
\begin{array}{ll}
S \dermp^{\mcR} & \stmp{\{\prim(2), \prim(3),\prim(5)\}}{X{=}6 \land B}{\{X\}}
\end{array}
\]
Hence, a single parallel step is sufficient to find all prime numbers.
} 
\end{example}

\subsection{Example Programs under Exhaustive Parallelism}
\label{sec:chrmpapp}

We examine different algorithms written in CHR and the effect of executing these programs in \chrmp, 
in particular with exhaustive parallelism to achieve maximum speed-up. 

\myparagraph{Filter Programs}
Programs that only consist of rules whose body is {\tt true} can be understood as filtering constraints.
They can obviously be executed in constant time with exhaustive parallelism, given enough processors.
The minimum and the prime program fall into this category. 
The {\tt msort} rule of merge sort leads to a linear number of exhaustively parallel steps. 
It can be rewritten to achieve constant-time complexity.
The experiments with the prime program using {massive parallelism} (see Section 8)
\cite{triossi2012compiling}
show an run-time improvement of about an order of magnitude over strong parallelism.

\myparagraph{SAT Solving}
The SAT formula is given as a tree of its sub-expressions. The tree nodes are of the form {\tt eq(Id,B)},
where {\tt Id} is a node identifier and {\tt B} is either a Boolean variable written {\tt v(X)}
or a Boolean operation ({\tt neg, and, or}) applied to identifiers. 
Additionally, a {\tt f(L,[])}
constraint is required in the initial state,
where {\tt L} is a list of all $n$ variables in the SAT formula. 
\begin{verbatim}
generate : f([X|Xs], A) <=> f(Xs,[true(X)|A]), f(Xs,[false(X)|A]).
assign   : f([],A) \ eq(T,v(X)) <=> true(X) in A | sat(T,A,true).
assign   : f([],A) \ eq(T,v(X)) <=> false(X) in A | sat(T,A,false).

sat(T1,A,S) \ eq(T,neg(T1)) <=> sat(T,A, neg S).
sat(T1,A,S1), sat(T2,A,S2) \ eq(T,and(T1,T2)) <=> sat(T,A, S1 and S2).
sat(T1,A,S1), sat(T2,A,S2) \ eq(T,or(T1,T2)) <=> sat(T,A, S1 or S2).
\end{verbatim}
The {\tt generate} rule generates, in $n$ parallel steps, $2^n$ {\tt f} constraints
representing all possible truth assignments to variables as a list in its second argument. 
In the next parallel step (using the {\tt assign} rules) all $n$ Boolean
variables in the given formula are assigned truth values for each
assignment, represented by {\tt sat} constraints. 

The remaining three rules
determine the truth values of all sub-expressions of the formula bottom-up. In each
parallel step the truth values of sub-expressions at a certain height of the tree are
concurrently computed for all possible assignments of variables. Therefore, the
number of parallel steps in this phase is bound by the depth of the formula. 

A formula is in 3-DNF normal form if it is in disjunctive normal form (a disjunction of conjunctions of literals) 
and each clause contains at most $3$ literals.
Because of its bounded depth, a SAT problem given in 3-DNF normal form with $n$ variables can be
solved in linear time in $n$ with this program under exhaustive parallelism, independent of the size of the formula.

\subsection{Properties: Soundness under Deletion-Acyclicity}

Soundness of \chrmp\ is not always possible as the
following example shows.
\begin{example}
{\rm
Consider the following rule that removes one of two differing constraints,
\begin{verbatim}
c(N) \ c(M) <=> N=\=M | true.
\end{verbatim}
and the goal {\tt c(1), c(2)}.  
There are two competing rule instances for application: one matches the
two constraints in the given order, the other in reversed order.
So if we apply both rules simultaneously under exhaustive parallelism, both constraints will be (incorrectly) removed.
} 
\end{example}
In general, computations that allow for mutual removal of constraints are not sound in \chrmp.
Soundness requires that the programs are
{deletion-acyclic}, effectively ruling out mutual removal. 
A {\em deletion dependency pair} $(c,d)$ means the kept constraint~$c$ is required to remove constraint~$d$ in a rule of the program.
This is the case if $c$ as an instance of a kept constraint and $d$ is an instance of a removed constraint in the head of the rule.
\begin{definition}[Deletion Dependency,
Deletion-Acyclic]\label{def:deletion_acyclic} 
{\rm
Given a \chrmp\ state $S = \stmod{\bbG}{\bbB}{\bbV}$.
Then \emph{deletion
dependency}~$\mcD(S)$ is a binary relation such that $(c,d) \in
\mcD$ if and only if there exist
$(H_1,H_2,B_c,B_b,\bbB')$ $\in \mcR(S)$ and $c' \in H_1, d' \in H_2$ such that $c'
\land \bbB' \rightarrow c$ and $d' \land \bbB' \rightarrow d$.

A \chrmp\ program~\mcP\ is \emph{deletion-acyclic} if and only if for all
$S$ such that $S \dermp^\mcR T$ the transitive closure~$\mcD(S)^+$ is irreflexive.
} 
\end{definition}

In a deletion-acyclic program, we can simulate the \chrmp\ computation steps by a sequence of sequential rule applications
in multiset semantics, provided we initially have enough copies of the user-defined constraints and can remove them when needed.
The latter is accomplished by so-called {\em set-rules} of the form 
\begin{verbatim}
set-rule: c(X1,...Xn) \ c(X1,...Xn) <=> true.
\end{verbatim}
for each CHR constraint {\tt c/n} in the given program. These rules remove multiple occurrences of the same constraint.

The following soundness theorem requires a deletion-acyclic program and set-rules \cite{raiser_fru_parallel_wlp10}.
Let $\der$ be a sequential transition in a suitable variant of the usual multiset CHR semantics.
\begin{theorem}[Soundness] \label{thm:soundness}
{\rm
Let $\mcP$ be a deletion-acyclic
\chrmp\ program and $\mcP'$ be the CHR program~$\mcP$ extended with 
{set-rules}. 
If $S = \stmod{\bbG}{\bbB}{\bbV} \dermp_\mcP T$,
then there exists a multiset~$\bbG'$ with $c \in \bbG'
\Leftrightarrow c \in \bbG$ such that 
$S' = \stmod{\bbG'}{\bbB}{\bbV} \der^{*}_{\mcP'} T'$, 
where 
$c \in T' \Leftrightarrow c \in T$.
} 
\end{theorem}

\begin{example}
{\rm
Consider the initial goal {\tt a} and the program
\begin{verbatim}
a <=> b,c.           a <=> b,d.        b,c,d <=> true.
\end{verbatim}
Exhaustive parallelism leads to the set-based computation\\
\medskip
{\tt a} $\dermp$ {\tt b,c,d} $\dermp$ true.\\
\medskip
The sequential correspondence in the multiset CHR program extended with set-rules is\\
\medskip
{\tt a,a} $\der$ {\tt b,b,c,d} $\der$ {\tt b,c,d} $\der$ true.
} 
\end{example}

The example can also be used to show that {\em Serializability} in general does not hold for massively-parallel set-based CHR. There is not sequential computation in \chrmp\ that can simulate the exhaustively parallel computation, since the first rule application will remove {\tt a}, so either {\tt b,c} or {\tt b,d} can be produced sequentially, but not their union.
Similarly, {\em monotonicity} does not hold.

\section{Parallel Hardware Implementations of CHR}

The work reported in \cite{triossi2012compiling,triossi_phd11} 
investigates the compilation of CHR to specialized hardware. 
The implementation follows the standard scheme for translating CHR into procedural languages.
The compiler translates the CHR code into the low-level hardware description language VHDL, 
which in turn creates the necessary hardware using Field Programmable Gate Array (FPGA) technology.
FPGA is a hardware consisting of programmable multiple arrays of logic gates. 
We also implement a hybrid CHR system consisting of a software component running a CHR system for sequential execution, 
coupled with hardware for parallel execution of dedicated rules in the program. 
The resulting hardware system is typically an order of magnitude faster than the fastest software implementation of CHR (in C).

\subsection{Basic Compilation of CHR to Procedural Languages}

As preliminaries, we give the basic implementation scheme for Ground CHR in procedural languages like C and Java, but also VHDL. 
This translation scheme applies throughout this section.
In Ground CHR, we do not need to wake-up constraints, because all variables are ground at run-time.
A CHR rule can be translated into a procedure using the following simple scheme:

\medskip
\noindent \verb! procedure(!\textit{kept\_head\_constraints, removed\_head\_constraints}\verb!) {!\\
\verb!    if (!\textit{head constraints not marked removed}
\verb! && !\textit{head matching}\verb! && !\textit{guard check}\verb!)!\\
\verb!    then {!\textit{remove removed\_head\_constraints}\verb!; !\textit{execute body constraints}\verb!;} !\\
\verb! }!

\medskip
\noindent The parameter list references the head constraints to be matched to the rule. 
In the procedure, we first check that the constraints have not been marked as removed. 
Then head matching is explicitly performed and then the guard is checked.
If all successful, one removes the removed head constraints, executes the built-in constraints and then adds the body CHR constraints. 
Added constraints may overwrite removed head constraints for efficiency. 
Constraints that are removed and not overwritten are marked as deleted.
Such a rule procedure is executed on every possible combination of constraints from the store, typically through a nested loop
(that can be parallelized).
This basic translation scheme corresponds to the abstract semantics, since it does not distinguish between active and suspended CHR constraints.
It needs to be refined to be practical \cite{vanweert_lazy_evaluation_tkde10}.

\subsection{Compiling CHR to Parallel Hardware}

Our compiler translates the CHR code into the low-level hardware description language VHDL, 
which in turn creates the necessary hardware using FPGAs.
The architecture of FPGA hardware is basically divided into three parts: the internal computational units called configuration logic blocks, the Input/Output (I/O) blocks that are responsible for the communication with all the other hardware resources outside the chip, and the programmable interconnections among the blocks called routing channels. In addition, there can be complex hardware blocks designed to perform higher-level functions (such as adders and multipliers), or embedded memories, as well as logic blocks that implement decoders or mathematical functions.

\myparagraph{CHR Fragment with Non-Increasing Rules}
We assume Ground CHR. 
Since the hardware resources can only be allocated at compile time,
we need to know the largest number of constraints that can occur
in the constraint store during the computation.
In {\em non-increasing rules}, the number of body CHR constraints added is not greater than the number of head constraints removed. Thus the number of constraints in the initial goal provides an upper bound on the number of constraints during the computation.
Hence we only allow for non-increasing simpagation rules.

\myparagraph{CHR Compilation Hardware Components}
A Program Hardware Block (PHB) is a collection
of Rule Hardware Blocks (RHBs), each corresponding to a rule of the
CHR program. A Combinatorial Switch (CS) assigns the constraints to the PHBs.
In more detail:
\begin{description}
\item[Rule Hardware Block (RHB)]
In VHDL the rule is translated into a single clocked process following the transformation scheme described above.
Here, the parameters are input signals for each argument of the head constraints. 
Each signal is associated with a validity signal to indicate if the associated constraint has been removed.
A concrete example is given below.

\item[Program Hardware Block (PHB)]
The PHB makes sure that the RHBs keep applying themselves until the result remains unchanged for two consecutive clock cycles. Each rule is executed by one or more parallel processes that fire synchronously every clock cycle. 
The initial goal is directly placed in the constraint store from which several instances
of the PHB concurrently retrieve the constraints.

\item[Combinatorial Switch (CS)]
The CS sorts, partitions and assigns the constraints to the PHBs, ensuring that the entire constraint store gets exposed to the rule firing hardware.
It acts as a synchronization barrier, allowing the faster PHBs to wait for the slower ones, then communicating the results between the blocks. It also reassigns the input signals to make sure that all constraint combinations have been exposed to the rule head matching.
\end{description}

\myparagraph{Strong Parallelism with Overlap}
For a given kept constraint, multiple RHBs are used to try rules with all possible partner constraints. 
For the case of simpagation rules with one kept and one removed constraint,
we introduce a hardware block that consists of a circular shift register which contains all the initial goal constraints. The first register cell contains the kept constraint and it is connected to the first input of all the RHBs, the rest of the register cells contain the potential partner constraints and are each connected to the second input of one RHB. Every time the PHBs terminate their execution, the new added constraints replace the removed ones. They registers shift until a non-removed constraint is encountered.
\begin{example}
{\rm
Consider the rule for the greatest common divisor:
\begin{verbatim}
         r :  gcd(N) \ gcd(M) <=> M>=N | gcd(M-N).
\end{verbatim}
In Figure \ref{fig:VHDLgcd}
we give an excerpt of the VHDL code produced for the above rule.
There are two processes executed in parallel, 
one for each matching order,
that correspond to two RHBs called 
{\tt r\_1} and {\tt r\_2}. 
The input parameters {\tt gcd1} and {\tt gcd2} are byte signals holding the numbers.   
{\tt valid1s} and {\tt valid2s} are bit signals.   
They are set to {\tt 0} if the associated constraint is removed.
The shared variable {\tt flag} is a bit. 
It is used to control the application of the two processes.

\begin{figure}[htb]
\rule{\textwidth}{0.5pt}
\begin{verbatim}
  r_1: process (..., gcd1s, gcd2s, valid1s, valid2s)
  begin  
    if ...      % checking and setting flags and parameters
      if (valid1s=’1’ and valid2s=’1’) then
        if gcd2s>=gcd1s then
          gcd2s <= gcd2s-gcd1s;
          flag := ’1’;                         
        else
          flag := ’0’;
        end if;
      end if;
    end if;
  end process r_1;

  r_2: process (..., gcd1s, gcd2s, valid1s, valid2s)
  begin 
    if ...      % checking and setting flags and parameters
      if (valid1s=’1’ and valid2s=’1’) then
        if flag=’0’ then                        
          if gcd1s>=gcd2s then
            gcd1s <= gcd1s-gcd2s;
          end if;
        end if;
      end if;
    end if;
  end process r_2;
\end{verbatim}

\rule{\textwidth}{0.5pt}
\caption{Excerpt of VHDL Code for GCD Rule}	
\label{fig:VHDLgcd}
\end{figure}

}
\end{example}

\myparagraph{Massive Parallelism} 
The set-based semantics \chrmp\ (see Section 7) allows multiple simultaneous removals of the same constraint.
Our implementation eliminates the conflicts in the constraint removals by allowing different rule instances to work concurrently on distinct copies of the constraints.
We provide all possible combinations of constraints to distinct parallel PHB instances in a single step. So the same constraint will be fed to several PHBs. Valid constraints are collected. A constraint is valid if no PHB has removed it. This is realized in hardware by AND gates.
The improvement
due to massive parallelism is about an order of magnitude
for goals with a low number of constraints and it decreases with
higher numbers of constraints. This is due to reaching the
physical bounds of the hardware. 

\myparagraph{Experimental Results}
A few experiments were performed including the programs for Minimum, Prime Numbers, GCD, Merge Sort, Shortest-Path and Preflow-Push\cite{triossi2012compiling,triossi_phd11}.
Unfortunately, no tables with concrete performance numbers are given, just log-scale diagrams. From them we can see the following.
The FPGA implementations of CHR are at least one order of magnitude faster than the fastest software implementations of CHR.
In the experiments, Shortest-Path and Preflow-Push showed a consistent parallel speed-up.
Strong parallelism improves the performance, and massive parallelism improves it further by up to an order of magnitude for the Prime example.
In the examples, the code produced by the CHR-to-FPGA compiler is slower but within the same order of magnitude as handcrafted VHDL code.

\myparagraph{Translation into C++ for CUDA GPU}
Graphical Processing Units (GPUs) consist of hundreds of small cores to provide massive parallelism. Similar to the work on parallel CHR FPGA hardware, the preliminary work in \cite{zaki_parallel_gpu_chr12} transforms {\em non-increasing Ground CHR rules} to C++ with CUDA in order to use a GPU to fire the rules on all combinations of constraints. As proof of concept, the scheme was encoded by hand for some typical CHR examples. No experiments are reported.
The constraint store is implemented as an array of fixed length consisting of the structures that represent CHR constraints. 
A CHR rule can be translated into a function in C++ using the basic procedural translation scheme.
The rule is executed on every possible combination of constraints using nested for-loops. 
Finally, the code is rewritten for the CUDA library.
The outer for-loop is parallelized for the thread pools of the GPU.

\section{Distribution in CHR}\label{chrd}

Before we introduce a full-fledged distributed refined semantics for CHR and its implementation, 
we set the stage by describing a distributed but sequential implementation of set-based CHR.
This system is successfully employed in a verification system for concurrent software.
Both semantics work with a syntactic subset of CHR where head constraints in rules must share variables 
in specific ways to enable locality of computations. 
Both semantics feature propagation rules, but they use different mechanisms to avoid 
their repeated re-application.

\subsection{Distributed Set-Based Goal Stores in \chrd}

\chrd\ \cite{sarnastarosta_ramakrishnan_chrd_padl07}
is an implementation of a sequential {\em set-based} {\em refined} semantics for CHR
with propagation rules.
\chrd\ features a distributed constraint store.

\myparagraph{Termination of Propagation Rules}
There are basically two ways to avoid repeated application of
propagation rules: Either they are not applied a second time to the same
constraints or they do not add the same constraints a second time.
Since we can remove constraints in CHR, usually the first option is chosen:
we store the sequence of CHR constraint identifiers to which a propagation rule has been applied.
It can be garbage-collected if one of the constraints is removed.
This information is called a {\em propagation history}. 
\chrd\ replaces the check on the propagation history by an {\em occurrence check} on the constraint store. 
This can be justified by the set-based semantics.

\myparagraph{Set-Based Refined Semantics}
Our {set-based semantics} closely follows the standard {refined semantics} \cite{duck_stuck_garc_holz_refined_op_sem_iclp04}. 
The essential differences are as follows:
\begin{itemize}
\item The propagation history is dropped from the states.

\item There is an additional transition to ensure a set-based semantics. 
It removes a constraint from the goal store before its activation, if it is already in the constraint store. 

\item 
There are additional transitions to avoid immediate re-application of a propagation rule.
In the first transition, all head matching substitutions where the active constraint is kept are computed at once and all corresponding rule instances are added to the goal store. These rule instances are called {\em conditional activation events}.

\item When a conditional activation event is processed, it is checked if the matching head constraints are still in the constraint store. If not, a second transition removes the event from the goal store. Otherwise, a third transition applies the rule instance by adding its body constraints to the goal store.

\end{itemize}
The semantics does not model the distribution of the \chrd\ constraint store.

Our set-based semantics is not always equivalent to the standard refined semantics. In the semantics a propagation rule may fire again on a constraint that has been re-activated (woken). In the refined multiset semantics, it will not be fired again.
So a CHR program may not terminate with the set-based semantics, but with the refined semantics.

\myparagraph{Distributed Local Constraint Stores by Variable Indexing}
Finding the partner constraints in head matching efficiently is crucial for the performance of a CHR system.
If variables are shared among head constraints, we can use the corresponding arguments of the constraints for {\em indexing}.
If the argument is an unbound variable at run-time, we store (a pointer to) the constraint as attribute of that variable.
If the argument becomes bound (or even ground) at run-time, the constraint can be accessed from a hash table instead.

A conjunction of constraints is {direct-indexed (connected)} if all subsets of constraints share variables with the remaining constraints. 
In other words, it is not possible to split the constraints in two parts that do not share a variable.
\begin{definition}
{\rm The {\em matching graph} of a set $C$ of constraints is a labeled undirected graph $G = (V, E)$ where $V = C$, and $E$ is the smallest set such that $\forall c_1, c_2 \in V, vars(c_1) \cap vars(c_2) \neq \{\} \rightarrow (c_1, c_2) \in E$
where $vars(c)$ returns the set of variables in a constraint $c$.
A rule R in a CHR program is said to be {\em direct-indexed (connected)} if the matching graph for its head constraints is connected.
A CHR program is direct-indexed if all its rule heads are direct-indexed. 
}
\end{definition}
Clearly, head matching is significantly improved for direct-indexed programs.
Instead of combinatorial search for matching partner constraints, constant-time lookups are possible with indexing.
\chrd\ requires direct-indexed programs that only index on unbound variables.
This permits the constraint store to be represented in a distributed fashion as a network of constraints on variables.

Any CHR program can be trivially translated to a direct-indexed program.
We just have to add an argument to each CHR constraint that always contains the same shared variable.
For example, the direct-index rule for minimum is:
\begin{verbatim}
min(X,N) \ min(X,M) <=> N=<M | true.
\end{verbatim}
With the help of the new variable, we can distinguish between different minima.
In general, this technique can be used to localize computations.

\myparagraph{Implementation and Experimental Results}
We have an implementation of ground \chrd\ in the Datalog fragment of Prolog, where terms are constants only.
Our implementation has been integrated into XSB, a Prolog programming system with {tabling}.
It can be obtained online with a free download from \url{http://xsb.sourceforge.net}.
\chrd\ performs significantly better on programs using {tabling}, 
and shows comparable results on non-tabled benchmarks. 
This indicates that constraint store occurrence checks can be done as efficiently as propagation history checks while avoiding the maintenance of a propagation history.

\myparagraph{Verification of Multi-Threaded Applications}
The paper \cite{ss_stirewalt_dillon_checking_deadlock_ijseke07} describes an approach for checking for deadlocks in multi-threaded applications based on the concurrency framework SynchroniZation Units MOdel (Szumo) 
\cite{Sarna-Starosta:2008:CAS:1521523}.
The framework associates each thread with a synchronization contract that governs how it must synchronize with other threads. At run-time, schedules are derived by negotiating contracts among threads. 

The Szumo system includes a constraint solver written in \chrd\ encoding the synchronization semantics of thread negotiation. 
The verification system performs a reachability analysis: it constructs execution paths incrementally until either a deadlock is detected or further extending the path would violate a synchronization contract. 

With Szumo, we analyzed an implementation of the {dining philosophers problem}, where no deadlock was found.
We verified the in-order message delivery property of an $n$-place FIFO buffer. 
We also analyzed Fischer’s protocol, a mutual-exclusion protocol that is often used to benchmark real-time verification tools. There we employed \chrd\ to specify a solver for the clock constraints.

\subsection{Distributed Parallel \chre\ and its Syntax}

The paper \cite{lam2013decentralized} introduces a decentralized distributed execution model consisting of an ensemble of computing entities, each with its own local constraint store and each capable of communicating with its neighbors:
in \chre, 
rules are executed at one location and can access the constraint stores of its immediate neighbors.
We have developed a prototype implementation of \chre\ in Python with MPI (Message Passing Interface) as a proof of concept and demonstrated its scalability in distributed execution. 
It is available online for free download at \url{https://github.com/sllam/msre-py}.

\myparagraph{Syntax of \chre}
We assume {Ground CHR}. 
\chre\ introduces locations.
\begin{definition}
{\rm
All user-defined constraints in a program must be explicitly localized.
A {\em location} $l$ is a term 
(typically an unbound variable or constant) 
that annotates a CHR constraint $c$, written as $[l]c$. 
A location $l$ is {\em directly connected} to a location $l'$ if there is a constraint $[l]c$ at location $l$ 
such that $l \in vars(c)$.
}
\end{definition}
We are interested in rules that can read data from up to $n$ of their immediate neighbors, but can write to arbitrary neighbors. We therefore define {\em $n$-neighbor restricted (star-shaped) rules} (which are a subclass of direct-indexed rules introduced in \chrd).
The rule head refers to directly connected locations in a star topology. At the center of the star is the {\em primary location}.
\begin{definition}
{\rm
A CHR rule with $n+1$ head constraints is {\em $n$-neighbor restricted (star-shaped)} if and only if there is a dedicated location called {\em primary location} and $n$ {|em neighbor locations} in the rule head satisfying the following conditions:
\begin{itemize}
\item The primary location is directly connected to each of its $n$ neighbor locations. 

\item If a variable is shared between constraints at different locations, it also must occur in the primary location.

\item Each constraint in the guard shares variables with at most one neighbor location.
\end{itemize}
}
\end{definition}
This definition ensures that computation can be structured and distributed by considering interactions between the primary location and each neighboring location separately.

\begin{example}
{\rm
This variant of the Floyd-Warshall algorithm computes all-pair shortest paths of a directed graph in a distributed manner. 
\begin{verbatim}
base  : [X]arc(Y,D) ==> [X]path(Y,D).
elim  : [X]path(Y,D1) \ [X]path(Y,D2) <=> D1<D2 | true.
trans : [X]arc(Y,D1), [Y]path(Z,D2) ==> X\=Z | [X]path(Z,D1+D2).
\end{verbatim}
We distinguish between arcs and paths.
{\tt [X]path(Y,D)} denotes a path of length {\tt D} from {\tt X} to {\tt Y}. 
The rules {\tt base} and {\tt elim} are $0$-neighbor restricted (local) rules because their left-hand sides involve constraints from exactly one location. 
Rule {\tt trans} is a $1$-neighbor restricted rule since its left-hand side involves {\tt X} and a neighbor {\tt Y}. 
We see that {\tt X} is the {primary location} of this rule because it refers to location {\tt Y} in an argument. 
} 
\end{example}

\subsection{Refined Semantics of \chre}

Before we discuss the refined semantics, we shortly mention the abstract semantics of \chre to introduce the basic principles.

\myparagraph{Abstract Distributed \chre\ Semantics for $n$-Neighbor Restricted Rules}
Each location has its own goal store.
Based on the standard abstract CHR semantics, we introduce abstract {\em ensemble} states, which are sets of local stores $G_k$ where $G$ is a goal and $k$ a unique location name.
In the adapted {\bf (Apply)} transition, each of the locations in an $n$-neighbor rule provides a partial match in their stores. If the matchings can be combined and if the guard holds, we add the rule body goals to their respective stores.
We show {\em soundness} with respect to the standard CHR abstract semantics, where locations are encoded as an additional argument to each CHR constraint.

\myparagraph{Refined Distributed \chre\ Semantics for $0$-Neighbor Restricted Rules}
We extend the standard CHR refined semantics to support decentralized incremental multiset matching for 0-neighbor restricted rules. 

\myparagraph{Localized States}
In \chre, an {\em ensemble} $\Omega$ is a set of localized states.
A {\em localized state} is a tuple $\langle \vec{{U}}; \vec{{G}}; \bar{{S}}; \bar{{H}} \rangle_k$, where
\begin{itemize}

\item the {\em Buffer} $\vec{{U}}$ is a queue of CHR constraints that have been sent to a location,

\item the {\em Goal Store (Execution Stack)} $\vec{{G}}$ is a stack of the constraints to be executed,

\item the {\em Constraint Store} $\bar{{S}}$ is a set of identified constraints to be matched,

\item 
the {\em Propagation History} $\bar{{H}}$ is a set of
sequences of identifiers of
constraints that matched the head constraints of a rule, 

\item the state is at {\em location} $k$.
\end{itemize}
To add a further level of refinement, an 
{\em active occurrenced CHR constraint}
$\act{c(\bar x)}{i}{j}$ is an identified constraint that is only allowed to match with the $j$-th occurrence
of the constraint predicate symbol $c$ in the head of a rule of a given CHR program $\mathcal{P}$.

To simplify the presentation of the semantics, we assume static locations: for all locations occurring in a computation, 
there is a localized state (possibly with empty components) in the ensemble.

\ELfig{fig:chre_semantics}{The Sequential Part of the Refined \chre\ Semantics for $0$-Neighbor Restricted Rules}{
{
\bda{c} 
   \ba{cc}

    \ELtlabel{\bf Flush} & 
    \ELmyirule{\vec{{U}} \neq \{\}}
    {\Omega, \langle \vec{{U}}; \{\}; \bar{{S}}; \bar{{H}} \rangle_k \mapsto 
     \Omega, \langle \{\}; \vec{{U}}; \bar{{S}}; \bar{{H}} \rangle_k} \\ \\

    \ELtlabel{\bf MoveLoc} & 
    \ELmyirule{}
    {\Omega, \langle \vec{{U}}; ({[k']c},\vec{{G}}); \bar{{S}}; \bar{{H}} \rangle_k,
     \langle \vec{{U}}; \vec{{G}}; \bar{{S}}; \bar{{H}} \rangle_{k'} 
     \mapsto 
     \Omega, \langle \vec{{U}}; \vec{{G}}; \bar{{S}}; \bar{{H}} \rangle_k,
     \langle (\vec{{U}},[c]); \vec{{G}}; \bar{{S}}; \bar{{H}} \rangle_{k'}} \\ \\

    \ELtlabel{\bf DropLoc} & 
    \ELmyirule{}
    {\Omega, \langle \vec{{U}}; ({[k]c},\vec{{G}}); \bar{{S}}; \bar{{H}} \rangle_k \mapsto 
     \Omega, \langle \vec{{U}}; (c,\vec{{G}}); \bar{{S}}; \bar{{H}} \rangle_k} \\ \\

    \ELtlabel{\bf Activate} & 
    \ELmyirule{d \mbox{ is a fresh identifier}}
    {\Omega, \langle \vec{{U}}; (c,\vec{{G}}); \bar{{S}}; \bar{{H}} \rangle_k \mapsto 
     \Omega, \langle \vec{{U}}; (c\#d:1,\vec{{G}}); (\bar{{S}},c\#d); \bar{{H}} \rangle_k} \\ \\

    \ELtlabel{\bf Remove} & 
    \ELmyirule{ \mbox{Variant of } (r~\atsign~[l]H_P' \backslash [l]H_S' \ELsimparrow C \mid B) \in {\cal P} \mbox{ such that  } \models \phi(C)  \ELsgap k=\phi(l)\\
                        \phi(H_P') = DropIds(H_P)  \ELsgap 
                        \phi(H_S') = \{c\} \cup DropIds(H_S)}
    {\Omega, \langle \vec{{U}}; (c\#d:i,\vec{{G}}); (\bar{{S}},H_P,H_S,c\#d); \bar{{H}} \rangle_k \mapsto 
     \Omega, \langle \vec{{U}}; (\phi(B),\vec{{G}}); (\bar{{S}},H_P); \bar{{H}} \rangle_k} \\ \\

    \ELtlabel{\bf Keep} & 
     \ELmyirule{ \mbox{Variant of } (r~\atsign~[l]H_P' \backslash [l]H_S' \ELsimparrow C \mid B) \in {\cal P} \mbox{ such that  } \models \phi(C) \ELsgap k=\phi(l) \\
                        \phi(H_S') = DropIds(H_S)  \ELsgap 
                        \phi(H_P') = \{c\} \cup DropIds(H_P) \ELsgap h = (r,Ids(H_P,H_S)), h \not\in \bar{{H}}}
   {\Omega, \langle \vec{{U}}; (c\#d:i,\vec{{G}}); (\bar{{S}},H_P,H_S,c\#d); \bar{{H}} \rangle_k \mapsto 
     \Omega, \langle \vec{{U}}; (\phi(B),c\#d:i,\vec{{G}}); (\bar{{S}},H_P,c\#d); (\bar{{H}},h) \rangle_k} \\ \\

    \ELtlabel{\bf Suspend} & 
    \ELmyirule{\mbox{(Remove) and (Keep) do not apply for }c\#d:i,\mbox{ occurrence } i \mbox{ exists}}
    {\Omega, \langle \vec{{U}}; (c\#d:i,\vec{{G}}); \bar{{S}}; \bar{{H}} \rangle_k \mapsto 
     \Omega, \langle \vec{{U}}; (c\#d:({i+1}),\vec{{G}}); \bar{{S}}; \bar{{H}} \rangle_k} \\ \\

    \ELtlabel{\bf Drop} & 
    \ELmyirule{i \mbox{ is not an occurrence in the program } \cal{P}}
    {\Omega, \langle \vec{{U}}; (c\#d:i,\vec{{G}}); \bar{{S}}; \bar{{H}} \rangle_k \mapsto 
     \Omega, \langle \vec{{U}}; \vec{{G}}; \bar{{S}}; \bar{{H}} \rangle_k} \\ \\

   \ea
\eda
} 
} 

\myparagraph{Localized Sequential Transitions}
Figure \ref{fig:chre_semantics} shows the sequential transitions for a single location.
\begin{itemize}
\item The {\bf (Flush)} transition step applies if the goal store is empty and the buffer is non-empty.
It moves all buffer constraints into the goal store. 
\end{itemize}

The transitions {\bf (DropLoc)} and {\bf (MoveLoc)} apply if the first constraint 
in the goal store of location $k$ is one for location $[k']c$.
They deliver constraint $[k']c$ to location $k'$. 
\begin{itemize}
\item The {\bf (MoveLoc)} transition applies if $k'$ is distinct from $k$ and there exists a location $k'$. 
It it strips the location $[k]$ away and sends constraint $c$ to the buffer of $k'$.

\item The {\bf (DropLoc)} transition applies if $k'$ is the same as $k$.
The location $[k]$ is dropped. 

\end{itemize}

The remaining transitions apply to a location as to a state in the standard refined semantics. 
Buffers are ignored and remain unchanged.
The transitions model the activation of a constraint, the application of rules to it, and its suspension if no more rule is applicable.
These transitions are as in the standard refined semantics of CHR, except that here we take care of locations and handle a propagation history.
\begin{itemize}
\item In the {\bf (Activate)} transition, a CHR constraint $c$ 
becomes active (with first occurrence $1$) and is also
introduced as identified constraint into the constraint store.

\item The {\bf (Remove)} transition applies a rule where the active constraint is removed.
There is a substitution $\theta$ under which constraints from the constraint store match the head of the rule and satisfy its guard (written $\models \theta \land G$).

The auxiliary function DropIds removes the identifiers from identified constraints.

\item The {\bf (Keep)} transition is like the {\bf (Remove)} transition except that
the active constraint $c$ matches a kept constraint and it is checked if the application of the resulting rule instance 
has not been recorded in the propagation history. If so, the active constraint is kept and remains active.
The propagation history is therefore updated. (It remains unchanged in all other transitions.)
The function $Ids$ returns the identifiers of identified constraints.

\item In the {\bf (Suspend)} transition, the active constraint cannot be matched
against its occurrence in the rule head. One proceeds to the
next occurrence in the rules of the program. 
This makes sure that rules are tried in the order given in the program.

\item The {\bf (Drop)} transition, if there is no more occurrence to try, removes the
active constraint the goal store, but it stays suspended in the constraint store.
\end{itemize}

\ELfig{fig:chre_par_semantics}{The Parallel Part of the Refined \chre\ Semantics for $0$-Neighbor Restricted Rules}{
{
\bda{c} 
   \ba{ccc}

    \ELtlabel{\bf Intro-Par} & &
    \ELmyirule{\Omega \mapsto \Omega'}
    {\Omega \parmapsto \Omega'} \\ \\

    \ELtlabel{\bf Parallel-Ensemble} & &
     \ELmyirule{(\Omega_1, \Omega_2) \parmapsto (\Omega'_1, \Omega_2) \ \ \ \ \ \ \ (\Omega_1, \Omega_2) \parmapsto (\Omega_1, \Omega'_2)}
    {(\Omega_1, \Omega_2) \parmapsto (\Omega'_1, \Omega'_2)} \\ \\

   \ea
\eda
} 
} 

\myparagraph{Localized Parallel Transitions}
Figure \ref{fig:chre_par_semantics} shows the parallel transitions. They are particularly simple.
As usual, the transition {\bf (Intro-Par)} says that any sequential transition is a parallel transition.
Transition {\bf (Parallel-Ensemble)} allows to combine two independent transitions on non-overlapping 
parts of the state (ensembles, i.e. sets of disjoint locations) into one parallel transition. 
This means that computation steps on different localized states can be executed in parallel.

\subsection{Properties: Monotonicity, Soundness and Serializability}
In the refined \chre\ semantics,
{monotonicity} holds with respect to locations, this means computations can be repeated in any larger context of more locations. 
{Serializability} holds in that every parallel \chre\ computation can be simulated using sequential \chre\ transitions.
We also prove soundness of the refined \chre\ semantics with respect to the abstract \chre\ semantics. 

We say that a \chre\ program is {\em locally quiescent (terminating)} if given a reachable state, we cannot have any infinite computation sequences that do not include the (Flush) transition. Hence local quiescence guarantees that each location will eventually process the constraints delivered to its buffer.

{Serializability and soundness} of the encoding holds for {quiescent} programs: 
computations between commit-free states of $0$-neighbor restricted encodings have a mapping to computations of the original $1$-neighbor restricted program.

The corresponding theorems and their detailed proofs can be found in the appendix of \cite{lam2013decentralized}.

\subsection{Encoding $1$- and $n$-Neighbor Rules in Local Rules}

We give an encoding of the more general $1$-neighbor restricted rules into local, i.e. $0$-neighbor restricted rules.
We can do the same for $n$-neighbor restricted rules. 
In this way, a programmer can use $n$-neighbor rules while the translation generates the necessary communication and synchronization between locations.
The encodings are a block-free variation of a two-phase commit consensus protocol between locations.

\myparagraph{Two-Phase-Commit Consensus Protocol}
The protocol consists of two phases:
\begin{itemize}
\item {\em Commit-Request Phase (Voting Phase).}
The coordinator process informs all the participating processes about the transaction and to vote either commit or abort. 
The processes vote.
\item {\em Commit Phase.}
If all processes voted commit, the coordinator performs its part of the transaction, otherwise aborts it. The coordinator notifies all processes. The processes then act or abort locally.
\end{itemize}
The standard protocol can block if a process waits for a reply. Not so in the variation we use.

\myparagraph{Encoding $1$-Neighbor Restricted Programs}
According to the following scheme, we translate each $1$-neighbor restricted rule of the form
\begin{verbatim}
r : [X]Px, [X]Px', [Y]Py \ [X]Sx, [Y]Sy <=> Gx,Gy | Body.
\end{verbatim}
In the head, {\tt Px} are the persistent constraints and {\tt Px'} are the non-persistent constraints.
Constraints are {\em persistent} if they are not removed by any rule in the program.
In the guard, {\tt Gx} contains only variables from location {\tt X}.
In the rule scheme below, 
{\tt XYs} contains all variables from the rule head, and {\tt Xs} only the variables from location x.
\begin{verbatim}
 % Commit-Request Phase
 % match and send request to neighbor location
request : [X]Px,[X]Sx ==> Gx | [Y]r_req(Xs).       
 % match and send commit to primary location          
vote : [Y]Py,[Y]Sy \ [Y]r_req(Xs) <=> Gy | [X]r_vcom(XYs). % if Sx non-e.
vote : [Y]Py,[Y]Sy,  [Y]r_req(Xs) ==> Gy | [X]r_vcom(XYs). % if Sx empty

 % Commit Phase
 % remove non-persistent constraints at primary location and send commit
commit :  [X]Px \ [X]Px',[X]Sx, [X]r_vcom(XYs) <=> [Y]r_commit(XYs).
 % remove at neighbor location, add non-persistent and body constraints 
act   : [Y]Py \ [Y]Sy, [Y]r_commit(XYs) <=> [X]Px', Body.
 % otherwise abort, re-introduce removed constraints at primary location
abort : [Y]r_commit(XYs) <=> [X]Px', [X]Sx.           
\end{verbatim}
The rule scheme uses different {\tt vote} rules depending on the emptiness of {\tt Sx}.
If {\tt Sx} is empty, it should be possible to remove several instances of {\tt Sy} with the same request.
Note that the rule scheme requires a refined semantics where rules are tried in the given order, because we have to make sure that rule {\tt act} is tried before the abort rule {\tt abort}.

The rule scheme implements an {\em asynchronous and optimistic consensus protocol} between two locations of the ensemble. It is asynchronous because neither primary nor neighbor location ever block or busy-wait for responses. Rather they communicate asynchronously via the protocol constraints, while potentially interleaving with other computations.
The temporary removal of non-persistent constraints in the rule scheme ensures that the protocol cannot be interfered with.
It is optimistic because non-protocol constraints are only removed after both locations have independently observed their part of the rule head instance.
It is possible that some protocol constraints are left if the transaction did not commit, but these can be garbage-collected.

We can generalize the above encoding to $n$-neighbor restricted rules.

\myparagraph{CoMingle}
This new programming language can be characterized as an extension of \chre\ for distributed logic programming \cite{lam2015,Cervesato2016}.
There is a prototype on the Android operating system for mobile devices, see \url{https://github.com/sllam/CoMingle}.
One application was built both using CoMingle and by writing traditional code: the former was about one tenth of the size of the latter without a noticeable performance penalty.

\section{Models of Concurrency in CHR}

Theoretical and practical models of concurrency have been encoded in CHR. 
Such an effective and declarative embedding holds many promises:
It makes theoretical models executable. 
It can serve as executable specification of the practical models.
One can toy with alternative design choices.
The implementations can be formally verified and analyzed using standard and novel CHR analysis techniques.
Last but not least it allows to compare different models on a common basis.

We will shortly introduce some common models of concurrency by their implementation in CHR:
Software Transactional Memory, Colored Petri Nets, Actors and Join-Calculus.
Typically soundness and completeness results will prove the correctness of these embeddings.

\subsection{Software Transactional Memory STM} \label{subsec:stmchr}

We have already seen the description of STM and its use to implement parallel CHR in Haskell in Section \ref{sec:impt-par-chr-ghc}. 
Now we do it the other way round. 
For the STM model, as a starting reference see \cite{Shavit1997}, 
for a high-level description see \cite{Guerraoui:2008:CTM:1345206.1345233}.
The paper \cite{sulzm_chu_lopstr08} gives a rule-based specification 
of Haskell's Software Transactional Memory in parallel CHR
which naturally supports the concurrent execution of transactions.

We classify CHR constraints once more into operation constraints and data constraints.
We assume CHR rules where the head contains exactly one operation constraint and 
the body contains at most one operation constraint. 

\myparagraph{Shared Memory Operations}
We first model shared memory and its associated read and write operations in CHR.
\begin{verbatim}
read  : cell(L,V1) \ read(L,V2) <=> V1=V2.
write : cell(L,V1), write(L,V2) <=> cell(L,V2).
\end{verbatim}
{\tt L} is a location identifier and {\tt V1} and {\tt V2} are values.
{\tt cell} is a data constraint, {\tt read} and {\tt write} are operation constraints.
The {\tt write} rule performs a destructive assignment to update the value of the cell.
With indexing and in-place constraint updates, the compiled rule can run in constant time.

\myparagraph{STM run-time manager in CHR}
The effects of an STM transaction are reads and writes to shared memory.  
The STM run-time must guarantee that all reads and writes within a transaction happen logically at once. In case transactions are optimistically executed in parallel the STM run-time must take care of any potential read/write conflicts. 
The STM run-time must ensure that in case of conflicts at least one transaction can successfully commit its updates whereas the other transaction is retried. 

To accomplish this behavior, we use for each transaction a read log and a write log.
Before we can commit the write log and actually update the memory cell, we first must validate that for each cell
whose value is stored in the read log, the actual value is still the same.

In Figure \ref{fig:STM}, we specify the STM manager via CHR rules. 
It has been slightly simplified in this survey.
Besides locations and values, we introduce an identifier for transactions {\tt T}.
The operation constraints are {\tt read} and {\tt write} and the protocol constraints are {\tt validate, commit} and {\tt rollback, retry}.
The data constraint {\tt CommitRight} acts as a token a committing transaction has to acquire in order to avoid concurrent writes.
The constraint {\tt validate} is issued at an end of the transaction if the {\tt CommitRight} is available.
Rules for rollback and retry of transactions are not shown here for space reasons.

\begin{figure}[htb]
\rule{\textwidth}{0.5pt}

\begin{verbatim}
 % Execution phase ---
 % Read from write or read log, create read log otherwise
r1 : WLog(t,l,v1) \ Read(t,l,v2) <=> v1=v2.
r2 : RLog(t,l,v1) \ Read(t,l,v2) <=> v1=v2.
r3 : Cell(l,v1) \ Read(t,l,v2) <=> v1=v2, RLog(t,l,v1). 

 % Write to write log, create write log otherwise
w1 : WLog(t,l,v1), Write(t,l,v2) <=> WLog(t,l,v2).
w2 : Write(t,l,v) <=> WLog(t,l,v).

 % Validation phase ---
 % Check and remove read log, rollback on read log conflict
v1 : Cell(l,v1), Validate(t) \ RLog(t,l,v2) <=> v1=v2 | True.
v2 : Cell(l,v1) \ Validate(t), RLog(t,l,v2) <=> v1=\=v2 | Rollback(t).

 % Start commit phase by acquiring CommitRight, otherwise rollback
s1 : CommitRight, Validate(t) <=> Commit(t).
s2 : Validate(t) <=> Rollback(t).

 % Commit phase ---
 % write update cells, then return CommitRight
c1 : Commit(t) \ Cell(l,v1), WLog(t,l,v2) <=> Cell(l,v2). 
c2 : Commit(t) <=> CommitRight.
\end{verbatim}
\rule{\textwidth}{0.5pt}
\caption{STM Run-Time Manager in CHR}	
\label{fig:STM}
\end{figure}

\myparagraph{Soundness and Correctness} 
Our implementation guarantees atomicity, isolation and optimistic concurrency. It is therefore sound.
It is correct: if a transaction commits successfully, the store reflects correctly all the reads/writes performed by that transaction.

\subsection{Colored Petri Nets CPN}

{\em Petri nets} are diagrammatic formalism to describe and reason about concurrent processes.
They consist of labelled {\em places} ($\bigcirc$) in which {\em tokens}
($\bullet$) reside. Tokens can move along {\em arcs} passing through {\em
transitions} ( \framebox(12,6){} ) from one place to another.
A transition may have several incoming arcs and several outgoing arcs. A transition
can only fire if all incoming arcs present a token. On firing, all incoming
tokens will be removed and a token will be presented on each outgoing arc.
{\em
Colored Petri Nets (CPN)} \cite{Jensen1987} significantly generalize Petri nets. 
Tokens are colored and places are typed by the colors they allow. 
Transitions can have conditions on tokens and equations that compute new tokens from old ones. 

The paper \cite{betz_petri_nets_chr07} shows that (Colored) Petri nets can easily be embedded into CHR. 
When CPNs are translated to CHR, color tokens are encoded as numbers. 
Place labels are mapped to CHR constraint symbols, tokens at a place to instances of CHR constraints, 
transitions and their arcs to simplification rules.
Incoming arc places form the rule head, outgoing arc
places form the rule body, and the transition conditions as well as equations form the rule guard.

\begin{example}\label{ex:philo_colored}
{\rm 
For simplicity, we consider the {dining philosophers problem} 
with just three philosophers as CPN
in Figure \ref{figure:philo_colored}.
Each philosopher (and fork) corresponds to
a colored token, given as a number from $0$ to $2$. 
Two philosophers $x$ and $y$ are
neighboring if $y = (x{+}1) ~\mathit{mod}~ 3$. 
Places are think, eat and fork, 
transitions are eat-think and and think-eat.

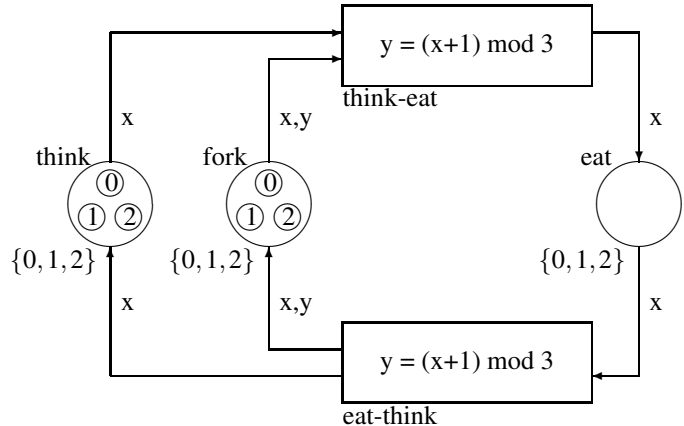
\begin{figure}[htb]
\rule{\textwidth}{0.5pt}
\begin{center}
{
\begin{picture}(250,200)(0,-100)   
	\put(40,0){\circle{30}}
	\put(100,0){\circle{30}}
	\put(240,0){\circle{30}}
	\put(12,15){\makebox{think}}
	\put(75,15){\makebox{fork}}
	\put(218,15){\makebox{eat}}
	\put(2,-24){\makebox{$\{0,1,2\}$}}
	\put(62,-24){\makebox{$\{0,1,2\}$}}
	\put(202,-24){\makebox{$\{0,1,2\}$}}
   
    %
    %
    \put(40,8){\circle{10}}
    \put(33,-5){\circle{10}}
    \put(47,-5){\circle{10}}
    \put(38,5){\makebox{0}}
    \put(31,-8){\makebox{1}}
    \put(45,-8){\makebox{2}}

    %
    %
    \put(100,8){\circle{10}}
    \put(93,-5){\circle{10}}
    \put(107,-5){\circle{10}}
    \put(98,5){\makebox{0}}
    \put(91,-8){\makebox{1}}
    \put(105,-8){\makebox{2}}
	
	%
	\put(128,-75){\framebox(94,30){y = (x+1) mod 3}}
	\put(128,45){\framebox(94,30){y = (x+1) mod 3}}
	\put(128,-83){\makebox{eat-think}}
	\put(128,37){\makebox{think-eat}}

    %
    \put(40,16){\line(0,1){49}}
    \put(40,65){\vector(1,0){88}}

    %
    \put(128,-65){\line(-1,0){88}}
    \put(40,-65){\vector(0,1){49}}

    %
    \put(100,16){\line(0,1){39}}
    \put(100,55){\vector(1,0){28}}
    
    %
    \put(100,-55){\vector(0,1){39}}
    \put(100,-55){\line(1,0){28}}

    %
    \put(222,65){\line(1,0){18}}
    \put(240,65){\vector(0,-1){49}}

    %
    \put(240,-16){\line(0,-1){49}}
    \put(240,-65){\vector(-1,0){18}}
    
    %
	%
	\put(44,30){\makebox{x}}
	\put(104,30){\makebox{x,y}}
	\put(244,30){\makebox{x}}
	
    %
	%
	\put(44,-40){\makebox{x}}
	\put(104,-40){\makebox{x,y}}
	\put(244,-40){\makebox{x}}

\end{picture}
} 
\end{center}
\rule{\textwidth}{0.5pt}
\caption{The Three Dining Philosophers Problem as Colored Petri Net}	
\label{figure:philo_colored}
\end{figure}

The CPN of Figure \ref{figure:philo_colored} translates into the following two CHR rules
\begin{verbatim}
think_eat : think(X), fork(X), fork(Y) <=> Y =:= (X+1) mod n | eat(X).
eat_think : eat(X) <=> Y =:= (X+1) mod n | think(X), fork(X), fork(Y).
\end{verbatim}
}
\end{example}

\myparagraph{Soundness and Completeness}
For both classical and Colored Petri nets, 
these correctness theorems are proven for the translation into CHR.

\subsection{Actor Model}

In the Actor Model \cite{Agha:1986:AMC:7929}, 
one coordinates concurrent computations by message passing. 
Actors communicate by sending and receiving messages. Sending is a non-blocking asynchronous operation. Each sent message is placed in the actors mailbox (a message queue). Messages are processed via receive clauses which perform pattern matching and guard checks. Receive clauses are tried in sequential order. The receive operation is blocking. If none of the receive clauses applies the actor suspend until a matching message is delivered.
Receive clauses are typically restricted to a single-headed message pattern. That is, each receive pattern matches at most one message. 

In \cite{sulz_lam_vanweert_actors_coordination08}, 
we extend the Actor Model 
      with receive clauses allowing for multi-headed message patterns. 
     Their semantics is inspired by their translation into CHR. 
We have implemented a prototype in Haskell 
\url{https://code.google.com/archive/p/haskellactor/}.

\begin{example}
{\rm In the Santa Clause problem, 
Santa sleeps until woken by either all of his nine reindeer or by three of his ten elves. If woken by the reindeer, he harnesses each of them to his sleigh, delivers toys and finally unharnesses them. If woken by three elves, he shows them into his study, consults with them on toys and finally shows them out. 
Here is a solution using the proposed multi-head extension:
\begin{verbatim}
santa sanActor =
 receive sanActor of
   Deer x1, Deer x2, ..., Deer x8, Deer x9 -> harness, deliver, unharness.
   Elf x1, Elf x2, Elf x3 -> enter_study, consult, leave_study.
\end{verbatim}
This straightforward solution avoids the clumsiness of explicitly counting deers and elves in the mailbox.
There is an obvious direct embedding of the matching receive clauses into CHR simplification rules.
} 
\end{example}

\myparagraph{Semantics of Actors with Multi-Headed Message Patterns}
We study two possible semantics for this extension, inspired by the standard {refined semantics} of CHR:
\begin{itemize}
\item The {\em first-match semantics} provides a conservative extension of the semantics of single-headed receive clauses. 
This semantics guarantees monotonicity:
any successful match remains valid if further messages arrive in the actor’s mailbox. 

\item The {\em rule-order-match semantics} guarantees that rule patterns are executed in textual order.  
In this semantics, newly arrived messages can invalidate earlier match choices.
\end{itemize}
It will depend on the application which semantics is the better choice.

\subsection{Join-Calculus and Join-Patterns}

In Join-Calculus \cite{Fournet2002}, 
concurrency is expressed via multi-headed declarative reaction rules that rewrite processes or events.
The (left-hand side of a) rule is called {\em join-pattern}.
They provide high-level coordination of concurrent processes. 
The thesis \cite{
lam_parallel_chr_11} extends join-patterns with guards
and describes a prototype implementation in parallel CHR compiled to Haskell, 
see \url{http://code.haskell.org/parallel-join}.

\myparagraph{Join-Calculus with Guarded Join-Patterns}
A concurrent {\em process (or event)}, say {\tt P}, has the form of a predicate.
A {\em reaction rule (join-pattern)} rewrites processes. 
We introduce {\em guards} into these rules:
$$\mbox{        Guarded Reaction Rule     } P_1,\ldots P_n \mbox{ if } Guard \Rightarrow P'_1,\ldots P'_m
$$
The Join-Calculus semantics is defined by a chemical abstract machine (CHAM). 
This model specifies transformations using a chemical reaction metaphor. 
The CHAM can be embedded in CHR, see Chapter 6 in \cite{fru_chr_book_2009}.

\begin{example}
{\rm
A print job is to be executed on any available printer where it fits. 
So print jobs have a size, and printers have a certain amount of free memory.
This behavior is captured by the following guarded reaction rule:
\begin{verbatim}
ReadyPrinter(p,m), Job(j,s) if m>s => SendJob(p,j)
\end{verbatim}
There is an obvious direct translation into CHR simplification rules.
} 
\end{example}

\myparagraph{Implementation and Experimental Results}
Standard CHR goal-based lazy matching is a suitable model for computing the triggering of join-patterns with guards: 
each process (CHR goal) essentially computes only its own rule head matches asynchronously and then proceeds immediately. 
We conducted experiments of our parallel Join-Calculus
implementation with examples for common parallel programming problems. 
They show consistent speed-up as we increase the number of processors.

\section{Discussion and Future Work}

We now present common topics and issues that we have identified as a result of this survey and that lead to research questions for future work.

\begin{table}[]
\centering
\begin{tabular}{|l||l|l|}
\hline
CHR Semantics & Syntactic Restriction                & Monotonicity Soundness Serializability     \\ \hline \hline
Abstract   Par.   & propagation rules do not terminate      & yes                                \\ \hline
Refined  Par.     & no propagation rules                     & yes                    \\ \hline
\chrmp        & no propagation rules                      & soundness for deletion-acyclic programs \\ \hline
\chrt         & ground data and operation constraints & yes               \\ \hline
\chrd         & direct-indexed rule heads                 & yes for ground confluent programs?                  \\ \hline
\chre         & ground star-shaped rule heads           & for quiescent programs   \\ \hline
\end{tabular}%
\caption{Syntactic Restrictions and Properties of CHR Parallel and Distributed Semantics}
\label{semoverview}
\end{table}

\myparagraph{Syntactic Fragments of CHR}
The parallel and distributed semantics surveyed are concerned with expressive Turing-complete fragments of CHR.
Their properties are summarized in Table \ref{semoverview}.
Except for the distributed semantics (\chrd\ and \chre) they do not allow for terminating propagation rules.
In the distributed semantics of \chrd\ and \chre\, one restricts rule heads to be sufficiently connected by shared variables,
requiring direct-indexed and $n$-neighbor (star-shaped) rules, respectively.
The former is no real restriction, the latter is.

Software implementations always presume Ground CHR (and so does \chrt).
Hardware implementations in addition rely on {\em non-size-increasing rules} which are still Turing complete.

Sometimes the notion of constraints is too abstract, and one differentiates between {\em data and operation constraints}.
Operation constraints update data constraints.
This dichotomy clarifies programs like Blocks World and Union-Find, 
is essential in the semantics of CHR with transactions (\chrt) and 
in the concurrency model of Software Transactional Memory when encoded in CHR.

All example programs in the survey and in general many other sequential CHR programs 
can still be run in parallel without modification,
since the syntactic restrictions are observed as they cover expressive subsets of CHR. 
However, changes are necessary if the program is not ground,
for parallel execution if the program contains propagation rules,
and for distributed execution if the rule heads are not sufficiently connected.
This need for program modifications weakens the promise of declarative parallelism, 
and therefore (semi-)automatic methods of program transformation should be investigated. 
Note that such transformations would be purely syntactical and do not require to come up with any scheduling for parallelism.

\myparagraph{Propagation Rules}
Surprisingly, while propagation rules seem perfect for parallelization (because they do not remove any constraints), 
they are currently only supported in distributed \chrd\ and \chre\ (see Table \ref{semoverview}).
(In the abstract parallel semantics, they are allowed, but do not terminate.)
On the other hand it seems possible to extend the refined parallel semantics with propagation rules, 
either using the {\em propagation history} of \chre\ or the {\em occurrence check} approach of \chrd\ to avoid their trivial non-termination.
The former seems to come with some implementation overhead, since the data structure needs to be updated in parallel.
The latter approach does not work in all cases, but it could be applicable to set-based semantics like \chrmp. 
As for a third possibility, in the literature on optimizing CHR implementations one can find program analyses that detect if 
propagation rules can be executed without any checks. 
Ground CHR is a good candidate for avoiding checks altogether, because constraints cannot be re-activated.

\myparagraph{Semantics Properties: Monotonicity, Serializability and Soundness}
These properties have been proven for all parallel CHR semantics based on multisets, 
for distributed \chre\ with the restriction to quiescent programs.
Surprisingly, these properties do not hold in general for the set-based semantics of distributed \chrd\ and massively-parallel \chrmp.
The papers on \chrd\ do not fully investigate these properties, while \chrmp\ is sound for {\em deletion-acyclic programs}.
Clearly, set-based semantics for CHR have to be studied more deeply.
There seems to be a mismatch between their elegance of the concept and its actual behavior.

\myparagraph{Program Analysis}
We should re-examine CHR program analysis for parallel and distributed CHR to see how they carry over.
{\em Termination} corresponds to {\em quiescence} in the concurrent context. 
There is a vast literature on (non-)termination and complexity analysis of CHR programs.
{\em Confluence} is an essential desirable property of sequential CHR programs. 
It already plays a role in parallel CHR for sound removal of transactions and seems trivial in exhaustively-parallel \chrmp. 
Confluence seems strongly related to soundness and serializability properties of concurrent CHR semantics.
Semi-automatic {\em completion} generates rules to make programs confluent. 
This method has been used in parallelizing the Union-Find algorithm and can be used for translating away CHR transactions.
When transactions are involved, confluence seems to avoid deadlocks.
We also think that the property of {\em deletion-acyclicity} of \chrmp\ has a broader application in rule-based systems. 
It seems related to confluence and we think can be expressed as a termination problem.

\myparagraph{Software and Hardware Implementations}
All software implementations surveyed are available online for free download, the links have been given.
The implementations cover parallel CHR, set-based \chrd\ and distributed \chre\ as well as CoMingle.
All implementations restrict themselves to the ground subset of CHR.
A full-fledged widely used stable implementation of parallel CHR is still missing. 
It could serve as a basis to foster further research and applications, as does the K.U. Leuven platform for sequential CHR.
With CoMingle, the situation seems better in the case of distributed CHR.
In any case, more evidence in the form of experimental results is needed to further confirm the promise of 
declarative concurrency made by CHR.

\myparagraph{Models of Concurrency in CHR}
Embedding models of concurrency in CHR is promising for understanding, analyzing and extending models, but still in its infancy.
It is appealing because of the {\em lingua franca} argument for CHR: different embeddings can be compared on its common basis and fertilize each other.
Conversely, the 
striking similarity of some models when encoded in CHR leads one to speculate about a generic concurrency model that is a suitable fragment of CHR which could then be mapped to many existing models, yielding a truly unified approach.

\section{Conclusions}
\label{sec:conclusion}

We have given an exhaustive survey of abstract and more refined semantics for parallel CHR as well as distributed CHR. Most of them have been proven correct. 
These semantics come with several implementations in both software and hardware. 
All software implementations are available online for free download.
We presented non-trivial classical example programs and promising experimental results showing parallel speed-up. 
Last but not least we reviewed concurrency models that have been encoded in CHR to get a better understanding of them and sometimes to extend them. 
Most of these embeddings have been proven correct, i.e. sound and complete. Some embeddings are available online.

In the discussion, we identified the following main topics for future work:
Including propagation rules into the parallel semantics and
providing program transformations into the expressive syntactic fragments for distributed CHR,
investigate set-based semantics and the deletion-acyclic programs,
provide a full-fledged implementation of parallel CHR,
apply the wealth of existing program analyses for sequential CHR to distributed and parallel CHR programs and the embedding of concurrency models,
and explore similarities of the concurrency models embedded in CHR as lingua franca to come up with unified models.

On a more general level, it should be investigated how the research surveyed here carries over to related languages like constraint logic programming ones and the other rule-based approaches that have been embedded in CHR.
Overall, the CHR research surveyed here should be related to more mainstream research in concurrency, parallelism and distribution.

\medskip
\myparagraph{Acknowledgements} We thank the anonymous referees for their helpful, detailed and demanding suggestions on how to improve this survey.

\bibliographystyle{alpha} 
\bibliography{devils,CHR2015,biblio,chr-biblio-jan-2017,chr-book,tfall2005,lamMAIN}

\label{lastpage}

\end{document}